\documentclass{aa} 
\usepackage{graphicx}
\usepackage{txfonts}
\usepackage{natbib}
\usepackage{longtable}
\usepackage{rotating}
\usepackage{lscape}
\usepackage{xcolor}
\usepackage{soul}
\usepackage[colorlinks,allcolors=blue]{hyperref}

\defcitealias{2001Carpenter}{CHS01}
\defcitealias{2015Rice}{RRW15}
\defcitealias{2017Roquette}{Paper I}
\defcitealias{gdw13}{GDW13}
\defcitealias{gdw15}{GDW15}
\defcitealias{hend2011+}{HSP11}

   \title{Near-infrared time-series photometry in the field of Cygnus OB2 association \thanks{The complete versions of Tables 1 to 5 and the light-curves of the stars analyzed in this study are available in electronic form at the CDS via anonymous ftp to \url{cdsarc.u-strasbg.fr (130.79.128.5)} \url{http://cdsarc.u-strasbg.fr/viz-bin/qcat?J/A+A/}}}

   \subtitle{II. Mapping the variability of candidate members}
   \author{J. Roquette
          \inst{1}\inst{2}
          \and
          S.H.P. Alencar\inst{2}
          \and
          J. Bouvier \inst{3}
          \and
         M. G. Guarcello\inst{4}
	  \and
	  Bo Reipurth\inst{5}
          }

   \institute{Department of Physics and Astronomy, University of Exeter, Physics Building, Stocker Road, Exeter, EX4 4QL, UK\\
              \email{jt574@exeter.ac.uk}
         \and
          Departamento de F\'isica - ICEx - UFMG, Av. Ant\^onio Carlos, 6627, 30270-901 Belo Horizonte, MG, Brazil\\
         \and
             Univ. Grenoble Alpes, CNRS, IPAG, F-38000 Grenoble, France \\
             \and
              INAF - Osservatorio Astronomico di Palermo, Piazza del Parlamento 1, I-90134, Palermo, Italy\\
	      \and
	      Institute for Astronomy, University of Hawaii at Manoa, 640
N. Aohoku Place, Hilo, HI 96720, USA\\
             }
   \date{\today}

  \abstract{We present the results of a J, H, and K photometric variability survey of the central 0.78 square degrees of the young OB association Cygnus OB2. We used data observed with the Wide-Field CAMera at the United Kingdom Infrared Telescope in 2007 (spanning 217 days) to investigate the light curves of 5083 low mass candidate members in the association and explore the occurrence and main characteristics of their near-infrared variability. We identified 2529 stars ($\sim$50$\%$ of the sample) with significant variability with time-scales ranging from days to months. We classified the variable stars into the following three groups according to their light curve morphology: periodic variability (1697 stars), occultation variability (124 stars), and other types of variability (726 stars). We verified that the disk-bearing stars in our sample are significantly more variable in the near-infrared than diskless stars, with a steep increase in the disk-fraction among stars with higher variability amplitude. We investigated the trajectories described by variable stars in the color-space and measured slopes for 335 stars describing linear trajectories. Based on the trajectories in the color-space, we inferred that the sample analyzed is composed of a mix of young stars presenting variability due to hot and cold spots, extinction by circumstellar material, and changes in the disk emission in the near-infrared. We contemplated using the use of near-infrared variability to identify disk-bearing stars and verified that 53.4$\%$ of the known disk-bearing stars in our sample could have been identified as such based solely on their variability. We present 18 newly identified disk-bearing stars and 14 eclipsing binary candidates among CygOB2 lower-mass members.} 


\begin{document} 
   \maketitle
\section{Introduction}

Early studies about stars in the premain sequence (PMS) phase \citep[e.g.,][]{1945Joy} have recognized photometric variability as one of the main characteristics of young stars. Exploring PMS variability in different regions of the electromagnetic spectrum provides insights into the physical processes at work in these systems. Optical and infrared (IR) variability studies as the Young Stellar Object (YSO) Variability Spitzer Space Telescope program \citep[YSOVAR ][]{MoralesCalderon2011,2014Rebull,2018Wolk} and the Coordinated Synoptic Investigation of NGC 2264 \citep[ e.g.,][]{2014Cody,2017Guarcello,2019Guarcello} have shown that most of the variability of a YSO is related to physical phenomena happening in the disk or due to the star-disk interaction. In disk-bearing PMS stars, the star-disk interaction and the accretion process causes variability due to the rotational modulation of hot spots at the stellar surface, which are produced at the base of the accretion column \citep[e.g.,][]{2016Sousa,2016Venuti}, instabilities in the accretion disk \citep[e.g.,][]{2007Bouvier} or in the accretion shock \citep[e.g.,][]{2008Koldoba}, variable accretion and the evolution of hot spots \citep[e.g.,][]{1988Vrba,1996Fernandez,2009Scholz}, and the obscuration of the PMS stellar photosphere by circumstellar material, as is the case of AA Tauri type stars \citep[e.g.,][]{2007Bouvier,2010Alencar,2014Fonseca}. Variability in PMS stars is not only restricted to those stars with disks. Rotational modulation by cool spots, caused by magnetic activity, can produce periodic optical and infrared photometric variability in both disk-bearing and diskless PMS stars \citep[e.g.,][]{2012Artemenko,2013Grankin}. 

In addition to probing the physical processes occurring during the PMS, variability can also be used to identify young stars. Knowing the typical variability characteristics of disk-bearing stars beforehand can lead to their identification in contrast to the low variability of field stars. This is especially powerful in the faint limits of optical and infrared surveys, where they are incomplete, and the young stars become blended with reddened background stars. In recent years, various variability surveys have used near-IR to unveil disk-bearing stars both in clusters and associations \citep[e.g.,][]{2012Rice,2013Wolk,2019Meng} as well as in the field \citep[e.g.,][]{2017Lucas}.
Recognizing the typical variability of disk-bearing stars requires determining the occurrence and prevalence of different variability mechanisms. Most of what we currently know about near-IR variability in the PMS is based on two studies by \citet[][hereafter \citetalias{2001Carpenter}, JHKs survey with 2MASS.]{2001Carpenter} and by \citet[][hereafter \citetalias{2015Rice}, JHK survey with the UKIRT]{2015Rice}, which altogether characterize the near-IR variability of $\sim$1800 young stars in the field of the Orion Nebula Cluster (ONC). In this paper, we contribute to the subject by characterizing the near-IR variability of $\sim$2500 PMS stars in the young and massive association Cygnus OB2.

Cygnus OB2 (hereafter CygOB2) is an OB association 1.33 kpc away from the Sun \citep{2015Kiminki} with estimated ages between 1 and 7 Myrs for its stellar population \citep{hanson2003,wright2010+}. CygOB2 harbors a rich and well studied population of massive stars counting with more than 150 OB stars \citep[e.g.,][]{knod2000,comeron2002+,2015Rauw,2015WrightMassive,2015Kiminki}, along with a low mass population of several thousand stars, which has been uncovered in the last decade by studies with \emph{Chandra} and \emph{Spitzer} \citep[][]{gdw13,Wright2014b+}. Previous variability studies in CygOB2 have focused on its OB population \citep[e.g.,][]{pigulski98+} and before this survey, the only variability survey partially concerning the low mass population was performed by \citet[][hereafter \citetalias{hend2011+}]{hend2011+} in two small 21$\farcm3\times23\farcm$3 fields where they observed in the R and I optical bands over two seasons of 19 and 18 observed nights and identified 121 variable stars. However their study only concerned the brightest low mass stars. 

In this paper, we investigated the near-IR variability of 5083 previously known low mass members in CygOB2. This work is complementary to the study presented in \citet[][hereafter \citetalias{2017Roquette}]{2017Roquette}, where we focused only on the stars with periodic behavior. Here, we look at all types of variability, regardless of periodicity. We found that 2529 of the known CygOB2 low mass members are variable to a significant level in the JHK bands, and we present their variability characteristics and major trends. Not only the results we present are an important step toward understanding the low mass population of CygOB2, and they also represent the largest single body of near-IR variable stars studied in the literature by almost doubling the number of such variables studied in the ONC by \citetalias{2001Carpenter} and \citetalias{2015Rice}.

This paper is organized as follows: We summarize the near-IR data used in Sec. \ref{sec:data} and present the methods applied for identifying and studying variable stars in Sec. \ref{sec:results}. We investigated variability in terms of Stetson Variability Index (Sec. \ref{sec:Stetson}), amplitudes (Sec. \ref{sec:amp}), light curve morphology (Sec. \ref{sec:morph}), and variability in the color-space (Sec. \ref{sec:corrcor} and \ref{sec:sec:comp}). We discuss our results in Sec. \ref{sec:discussion}. In Sec. \ref{sec:var} we show a comparison of the trends in color with the major physical mechanisms expected to produce variability in the near-IR.  In Sec. \ref{sec:sec:EB} we discuss the newly discovered eclipsing binaries in the region. Finally, in Sec. \ref{sec:diskcandidates}, we discuss the major variability characteristics of the known disk-bearing stars in our survey and use those to identify new disk-bearing stars in our sample. Finally, Sec. \ref{sec:conclusions} presents a summary of our results and concluding remarks.

\section{Analyzed sample}
\label{sec:data}

This paper is based on observations obtained in 2007 with the 3.8m United Kingdom Infra-Red Telescope (UKIRT), at Mauna Kea, Hawaii, equipped with the Wide Field Camera \citep[WFCAM,][]{wfcam}, programs U/07A/H16 and U/07B/H60 (P.I. Bo Reipurth). The complete dataset is composed of 115 observed nights in the J, H and K near-IR filters \citep{filtroukirt} and covers 217 days. The observations cover a field of 0.8 squared degrees, centered on $\alpha_{2000}=20^h33^m$, $\delta_{2000}=+41^o12'$. We refer to \citetalias{2017Roquette} for details on data reduction, calibration, and light curve production. Errors in the measurements used in this study are between 2$\%$ and 10$\%$ and their distributions are shown in Figure 5 of \citetalias{2017Roquette}.

\begin{figure*}
  \centering
  \includegraphics[width=17cm]{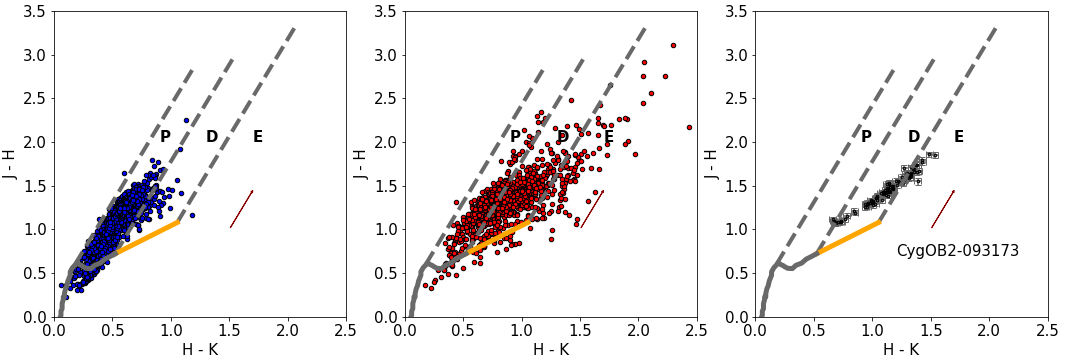}
  \caption{\label{fig:CCDm}Near-IR (JHK) color-color diagrams. Left and middle panels show the median colors of the 3811 diskless stars (blue dots) and the 1272 disk-bearing stars (red dots) analyzed in the present study. Right panel: shows the CygOB2-093173 variable star. Black dots show the color of the star in each observed night. The regions labeled as ``P'', ``D'' and ``E'' are the regions typically occupied by stars with normal photospheric colors, disk-bearing colors, and extreme near-IR excess respectively. The dark red arrow shows the effect of the interstellar reddening for the median visual extinction expected for stars in the CygOB2 association ($A_V=4.1^m$). The gray solid curve shows an empirical dwarf sequence by \citet{2007Kraus} and the orange line shows the \citet{1997Meyer} T Tauri locus. Dashed gray lines show extinction paths of $A_V=10^m$ for 0.6 M$_\odot$ and 0.08 M$_\odot$ stars, and the truncation point of the T Tauri locus.}
\end{figure*}

 The sample analyzed was composed of JHK light curves of previously known candidate members of CygOB2 that lay on the FOV of the survey (See \citetalias{2017Roquette} Figs. 3 and 4 for details). The list of known candidate members was built based on previous membership evaluation available in the literature, which includes disk-bearing stars identified by \citet[][hereafter \citetalias{gdw13}]{gdw13} and candidate members identified based on their X-ray properties as part of the
Chandra Cygnus OB2 Legacy Survey \citep[][hereafter \citetalias{gdw15}]{Wright2014b+,gdw15}. Out of 5083 objects analyzed, 7$\%$ (354) were in both X-ray sources, and disk-bearing candidate member lists, 18$\%$ (918) were only in the disk-bearing list, and 75$\%$ (3811) were only in the X-ray sources list. For this paper, we divided the analyzed sample in two: the sample of 1272 disk-bearing stars and the sample of 3811 members from X-ray not listed as disk-bearing stars (sample of diskless stars). Using GTC/OSIRIS \emph{riz} photometry from \citet{Guarcello2012+}, we inferred in \citetalias{2017Roquette} that the sample has masses mainly between 0.1 and 1.0 M$_\odot$.

Fig. \ref{fig:CCDm} shows a near-IR color-color diagram for the analyzed samples.
The left and middle panels show median colors for each star in the sample. The right panel shows individual data-points for the star CygOB2-093173, the most variable star in our sample. Three regions are marked (following \citet{2012Rice} and \citet{1996Itoh}): region ``P'' is typically occupied by stars with normal photospheric colors affected only by the interstellar extinction (Class III stars) and by reddened disk-bearing stars that show only a weak near-IR excess (Class II stars with only weak near-IR-excess); region ``D'' is typically occupied by stars with near-IR excess characteristic of disk-bearing stars (Class II stars); and region ``E'' is typically occupied by stars with extreme near-IR-excess possibly due to the emission of a circumstellar envelope (Class I stars). 
The reddening vector was built using the interstellar extinction law for the near-IR bands estimated by \citet{hanson2003} for the massive stars in CygOB2, transformed from 2MASS to UKIRT photometric system using the transformations proposed by \citet{filtroukirt}. 

The central panel in Fig. \ref{fig:CCDm} shows a significant number of disk-bearing stars identified by \citetalias{gdw13}, but located in the ``P'' region. The shorter wavelength in which a circumstellar disk produces IR-excess depends on the temperature distribution in the disk, and it may be the case that some disk-bearing stars do not present such excess in the near-IR but only at longer wavelengths. As \citetalias{gdw13} uses near- to far-IR to select disk-bearing stars, disk-bearing stars in the ``P'' region are likely stars that show an excess in the mid- and far-IR, but not in the near-IR. 

The left panel in Figure \ref{fig:CCDm} also shows a small number of diskless stars falling in the regions ``D'' and ``E''. These sources are identified as young by \citetalias{gdw15} and have near-IR excess typical of disk-bearing stars, but were missed by \citetalias{gdw13} presumably due to the different depths of the two surveys. Although \citetalias{gdw13} and \citetalias{gdw15} have similar completeness of about 0.9M$\odot$, the maximum depth of \citetalias{gdw13} is the one of the 
``Spitzer Legacy Survey of the Cygnus X region'' \citep{Beerer_2010}, $\sim$0.5M$\odot$, while \citetalias{gdw15} go as deep as $\sim$0.1M$\odot$. The possibility that these sources are indeed disk-bearing stars is discussed in Section \ref{sec:diskcandidates} in the context of near-IR variability as a disk-diagnosis.

\citetalias{gdw13} classifies the variable star CygOB2-093173 shown in the right panel of Fig. \ref{fig:CCDm} as Class II. However, because of the variability mechanism in action, the star describes a trajectory inside the color-color diagram, with a variability amplitude of $\sim$1 mag in the H-K color. It spends most of its observed time in the ``D’’ region, with a near-IR-excess compatible with a Class II classification, and part of its observed time in the ``E’’ region, with a near-IR-excess compatible with a Class I classification. Because of the ambiguities that the variability mechanism acting in disk-bearing stars can bring to the derivation of their evolutionary status based on single epoch observations, we do not adopt \citetalias{gdw13} evolutionary classes (Class I, Class II, or flat-spectrum) in our analysis. Instead, we considered that every star listed by them as a disk-bearing star can present variability caused by physical phenomena related to the disk or the star-disk interaction.

\section{Variability analysis}
\label{sec:results}

\subsection{Variable star selection}
\label{sec:Stetson}

\begin{figure*}
\centering
\includegraphics[width=5.5cm]{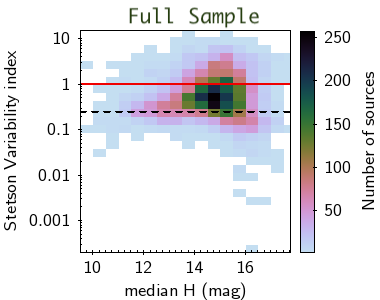}\includegraphics[width=5.5cm]{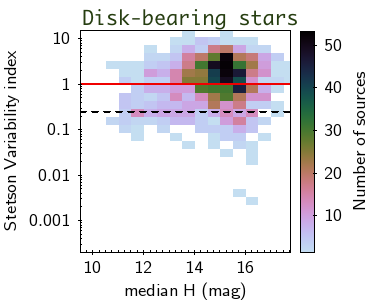}\includegraphics[width=5.5cm]{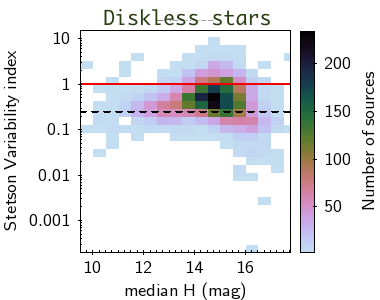}
\caption{\label{fig:stetmag}Distribution of Stetson variability index as a function of H median magnitude for the full sample of candidate members (left), disk-bearing stars (middle), and diskless stars (right). The red line shows $S=1$, and the dashed black line shows $S=0.25$.}
\end{figure*}

As in \citetalias{2017Roquette}, the initial selection of variable stars was done using the Stetson variability index \citep{1996Stetson}. The Stetson variability index ($S$) attributes $S\!\!\sim$0 to uncorrelated nonvariable stars and $S\!\!\geq$1 to significantly variable stars. Fig. \ref{fig:stetmag} shows $S$ as a function of median magnitudes in the H filter. The left, middle, and right plots show the full sample of candidate members, the disk-bearing stars, and the diskless stars, respectively. The mean, standard deviation and median value of $S$ for all the candidate members are $S(\mu,\sigma,\nu)$=0.9, 1.1,
0.5. The bulk of the $S$ distribution of disk-bearing stars is shifted toward larger values compared with the distribution of diskless stars. The former has $S(\mu,\sigma,\nu)$=1.8, 1.7, 1.3, and the latter has $S(\mu,\sigma,\nu)$=0.6, 0.6, 0.4. The difference between the diskless and disk-bearing samples is supported by a KS-test, resulting in a close to zero probability that the two distributions come from the same parent distribution, and corroborating the idea that disk-bearing stars are more variable in the near-IR than diskless stars.

Different authors adopt different Stetson variability index limits to select variable stars. For example, \citetalias{2001Carpenter} adopts $S\!\geq\!0.55$. Because in \citetalias{2017Roquette} we were interested in selecting low-amplitude periodic stars, we chose the limit $S\!=\!0.25$ for initial variable selection, which selects 4079 stars (80$\%$ of the full sample). In the present study we identified as significant variables stars with $S\!>\!1$ (sample $S_1$, 1280 sources). We also identified variables with $0.25\!\leq\!S<\!1.0$ if they presented at least one of the following characteristics. First if their light curves presented significant periodicity and they were among the 1154 sources discussed in Sec.\ref{sec:sec:per} and \citetalias{2017Roquette}. Second if their light curves presented occultation features: five sources discussed in Sec. \ref{sec:sec:ec}. Third if their light curves presented $\chi^2\geq$4.5, 4.5,5.7\footnote{The $\chi^2$-test is defined as
$\chi^2=\sum\limits^N_{i=1}\frac{(m_i-\bar{m})^2}{\sigma^2_i(N-1)}$, where $m_i$ is the $i$-th observation in a certain band, $\bar{m}$ is the weighted mean value, $\sigma_i$ is the photometric uncertainty for the $i$-th observation, and $N$ is the total number of observations. We calculated $\chi^2$ for each near-IR band and color. See \citet{2010Cody,2014Rebull,2017Sokolovsky} for examples of application of the $\chi^2$-test.} for the J, H and K filters respectively: 90 sources discussed in Sec. \ref{sec:sec:other}.

\begin{table*}
  \addtolength{\tabcolsep}{-2.8pt}
      \caption{\label{tab:NonVariable} General information for the nonvariable stars identified within this survey. The table shows the internal ID for each star, their coordinates, their ID in other studies: \citetalias{gdw13} and \citetalias{gdw15}, the number of valid points in their light curves in each band and their median magnitude, colors, and respective uncertainties. The full table is available online at the CDS.}

      \begin{tabular}{p{0.95cm}p{0.9cm}p{0.9cm}p{1.3cm}p{1.3cm}p{0.65cm}p{0.65cm}p{0.65cm}p{0.85cm}p{0.65cm}p{0.75cm}p{0.7cm}p{0.75cm}p{0.6cm}p{0.65cm}p{0.6cm}p{0.65cm}p{0.6cm}}
        \hline
        \hline
ID & GDW & GDW & RA & Dec & nJ & nH & nK & mJ  & eJ  & mH  & eH & mK & eK & mJH & eJH & mHK & eHK \\
&  13 & 15 &(deg) & (deg) & & & &  & & (mag) & &  & &  & (mag) & & \\
\hline
\hline
 000002 &  & 1044 & 308.0163 & 41.2944 & 103 & 101 & 101 & 16.80 & 0.03 & 15.56 & 0.03 & 14.99 & 0.02 & 1.25 & 0.04 & 0.56 & 0.04\\
  000005 &  & 1050 & 308.0186 & 41.2191 & 104 & 98 & 97 & 15.55 & 0.02 & 14.51 & 0.02 & 14.03 & 0.02 & 1.05 & 0.03 & 0.48 & 0.03\\
  000006 &  & 1056 & 308.0194 & 41.2128 & 104 & 97 & 100 & 16.57 & 0.03 & 15.35 & 0.02 & 14.80 & 0.02 & 1.21 & 0.04 & 0.56 & 0.03\\
  000009 &  & 1069 & 308.0215 & 41.2007 & 100 & 96 & 101 & 15.74 & 0.02 & 14.50 & 0.02 & 13.90 & 0.02 & 1.25 & 0.03 & 0.60 & 0.03\\
 \hline
 \end{tabular}
\end{table*}

  \begin{table*}
  \addtolength{\tabcolsep}{-2.8pt}
      \caption{\label{tab:data} General information for the Variable stars identified and studied in the present paper. The table shows the internal ID for each star, their coordinates, their ID in other studies:  \citetalias{gdw13} and \citetalias{gdw15}, the number of valid points in their light curves in each band and their median magnitude, colors, and respective uncertainties. The full table is available online at the CDS.}
      \begin{tabular}{p{0.95cm}p{0.9cm}p{0.9cm}p{1.3cm}p{1.3cm}p{0.65cm}p{0.65cm}p{0.65cm}p{0.85cm}p{0.65cm}p{0.75cm}p{0.7cm}p{0.75cm}p{0.6cm}p{0.65cm}p{0.6cm}p{0.65cm}p{0.6cm}}
        \hline
        \hline
ID & GDW & GDW & RA & Dec & nJ & nH & nK & mJ  & eJ  & mH  & eH & mK & eK & mJH & eJH & mHK & eHK \\
&  13 & 15 &(deg) & (deg) & & & &  & & (mag) & &  & &  & (mag) & & \\
\hline
   000001 &  & 1033 & 308.0141 & 41.3094 & 82 & 93 & 97 & 16.24 & 0.03 & 15.16 & 0.02 & 14.68 & 0.02 & 1.09 & 0.04 & 0.48 & 0.03\\
  000003 & 68396 & 1048 & 308.0173 & 41.2855 & 104 & 100 & 100 & 16.50 & 0.03 & 15.24 & 0.02 & 14.55 & 0.02 & 1.26 & 0.04 & 0.69 & 0.03\\
  000004 &  & 1047 & 308.0173 & 41.2385 & 103 & 99 & 101 & 16.61 & 0.03 & 15.35 & 0.02 & 14.78 & 0.02 & 1.26 & 0.04 & 0.57 & 0.03\\
  000007 &  & 1065 & 308.0212 & 41.2908 & 104 & 100 & 102 & 15.31 & 0.02 & 14.23 & 0.02 & 13.74 & 0.02 & 1.08 & 0.03 & 0.49 & 0.03\\
          \hline
      \end{tabular}

\end{table*}

\begin{table}
  \centering
      \caption{\label{tab:data2} 
Information about variability classification for the stars studied in the present paper. The Table shows the internal ID for each star; Stet gives their Stetson variability index. The flag Type indicates the morphology class attributed and can assume the values 1 for possibly periodic stars (Sec. \ref{sec:sec:per}), 2 for other types of variability (Sec. \ref{sec:sec:other}), and 3 for occultation variables (Sec. \ref{sec:sec:ec}). If a reliable period was measured, it is shown in the field Period; if the star was listed as disk-bearing by \citetalias{gdw13} the flag Disk is 1;  Comp is set if the star was classified as a compound variable in Sec. \ref{sec:sec:comp}: 1 - feature with distinct color, 2- variable IR-excess, 3- changing slope, 5- mix of 2- and 3-.; the flag Slope is set as 1 when the star had the slope of its trajectories in the color-magnitude and/or color-color diagram measured and is present in Table \ref{tab:slopes}. Asym is set if the light curve had an statistical asymmetry as in Sec. \ref{sec:symetry}: + if burst asymmetry, - if dimming asymmetry. The full table is available online at the CDS.}     
\addtolength{\tabcolsep}{-3.0pt}
      \begin{tabular}{p{1.2cm}p{0.9cm}p{0.9cm}p{0.9cm}p{0.7cm}p{0.7cm}p{0.8cm}p{0.7cm}}

        \hline
        \hline
\small{ID} & \small{Stet} & \small{Type} & \small{Period (d)} & \small{Disk} & \small{Comp} & \small{Slope} & \small{Asym}\\
\hline
  000001 &  0.6     &   2 &         &     & &   1 &\\      
  000003 &  1.28    &   1 &         & 1   & &    &\\      
  000004 &  0.46    &   1 & 8.35    &     & &    &\\      
  000007 &  0.52    &   1 & 2.82    &     & &    &\\      
\hline
      \end{tabular}
\end{table}

This procedure accounts for the selection of 1280 variable stars in the sample $S_1$, 1249 variable stars in the sample $S_2$, and the rejection of 1547 stars with $0.25\leq S<1.0$. For future reference, we show in Table \ref{tab:NonVariable} the general characteristics (coordinates, magnitudes, and colors) of the 2554 stars not considered as variable stars within this study. Table \ref{tab:data} presents the general characteristics for the 2529 sources considered as variable stars in our study. Table \ref{tab:data2} presents the variability flags assigned and described in the remaining of this section along with Stetson variability index and, for periodic stars, the measured period. In the rest of the paper, we only analyze the stars that we considered as variable stars.

\subsection{Variability amplitude}\label{sec:amp}

\begin{figure}
\centering
  \includegraphics[width=4.5cm]{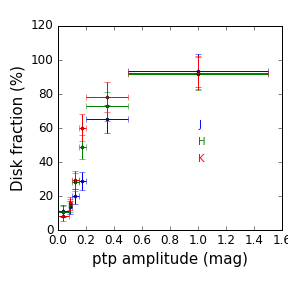}\includegraphics[width=4.5cm]{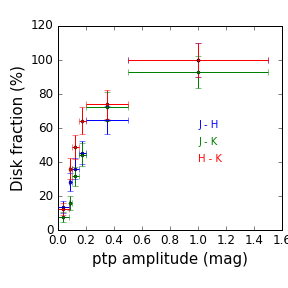}
  \caption{\label{fig:ampDfrac}Disk fraction as a function of peak to peak amplitude for each near-IR band (left) and color (right). In both plots, circles show the center of each ptp amplitude bin and horizontal bars indicate the size of each bin. Vertical bars show the standard errors of a Poisson counting in that bin.}
\end{figure}

Peak to peak (ptp) and root main square (rms) amplitudes were used to characterize the variability amplitudes in magnitude and color, and the values measured for each variable star are presented in Table \ref{tab:amp}. The differences between the photometric variability in disk-bearing and diskless stars are evident from a simple comparison between typical values for the two samples, and we verified that disk-bearing stars amplitudes are typically higher than diskless stars amplitudes. For example, the maximum, mean, and standard deviation of ptp amplitudes in the J band were 1.79, 0.27, and 0.22 mag for disk-bearing stars, and 0.94, 0.10, and 0.07 mag for diskless stars. For the H-K color, the maximum, mean, and standard deviation of ptp amplitudes were 0.97, 0.14, and 0.10  mag for disk-bearing stars and 0.41, 0.06, and 0.04 for diskless stars. Additionally, Fig. \ref{fig:ampDfrac} shows that there is a steep increase in the disk-fraction for larger amplitudes and that nearly all the stars with an amplitude larger than about 0.5 magnitudes are disk-bearing stars.

\begin{table*}
    \centering
  \addtolength{\tabcolsep}{-1.0pt}
      \caption{\label{tab:amp} Peak to peak and root mean square amplitudes in each observed band and color for each of the variable stars studied in the present paper. The full table is available online at the CDS.}
    \footnotesize{
      \begin{tabular}{p{2.15cm}|p{0.95cm}p{0.95cm}p{0.9cm}p{1.0cm}p{1.0cm}p{1.0cm}|p{0.9cm}p{0.9cm}p{1.0cm}p{1.0cm}p{1.0cm}p{1.0cm}}
        \hline
        \hline
 & \multicolumn{6}{c|}{ptp}  & \multicolumn{6}{c}{rms}  \\
        \hline
        ID &  J &H & K & J-H & J-K & H-K &  J &  H & K &  J-H & J-K &   H-K \\        
        \hline
\hline
CygOB2-00001 & 0.21 & 0.11 & 0.08 & 0.19 & 0.23 & 0.11 & 0.19 & 0.08 & 0.06 & 0.08 & 0.10 & 0.04\\
CygOB2-00003 & 0.16 & 0.14 & 0.14 & 0.07 & 0.11 & 0.08 & 0.11 & 0.10 & 0.10 & 0.03 & 0.04 & 0.03\\
CygOB2-00004 & 0.08 & 0.06 & 0.06 & 0.07 & 0.07 & 0.05 & 0.06 & 0.04 & 0.04 & 0.02 & 0.02 & 0.02\\
\hline
      \end{tabular}
      }
      \end{table*}

\subsection{Light curve morphology} 
\label{sec:morph}

\begin{figure}
  \includegraphics[height=9cm]{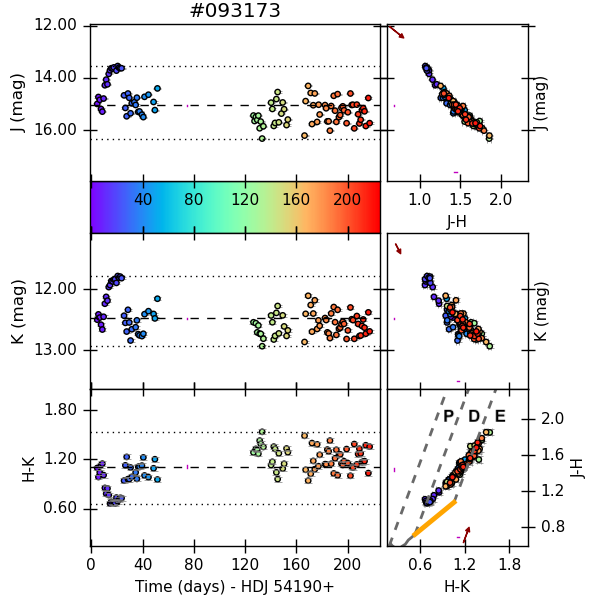}
  \caption{\label{fig:093173} Star CygOB2-093173 is a disk-bearing (Class II) star and an example of non periodic variable star, with $S=15.64$, and that presents reddening linear trajectories in the color-color and color-magnitude diagrams. \emph{Left}, from \emph{top} to \emph{bottom}: J light curve, a color bar showing the color attributed for each observation date, K light curve, and H-K color curve. The dotted lines show maximum and minimum values in each light curves, and dashed lines show the median values. \emph{Right}, from \emph{top} to \emph{bottom}: J vs. J-H color-magnitude diagram, K vs. H-K color-magnitude diagram, and color-color diagram with P, D, and E regions (discussed in Sec. \ref{sec:data}). The dark red arrows show the effect of an interstellar reddening of $A_\mathrm{V}=2^m$. The vertical and horizontal magenta line segments show the median error in each axis. In each plot, error bars are shown in gray. When visible and close to the position of the star, a gray curve shows a \citet{2007Kraus} empirical isochrone in the color-color diagram, and the orange line shows the T Tauri Locus. Each diagram has a range of 5$\sigma$ around the median color or magnitude for that star.}
\end{figure}

We visually inspected the light and color curves and classified the variable stars according to their morphology as shown in Fig. \ref{fig:093173}. The example in Fig. \ref{fig:093173} shows the disk-bearing star CygOB2-093173, whose color-color diagram presented in Fig. \ref{fig:CCDm}. We describe in the following sections the details of different types of variability morphology present in our sample.

\subsubsection{Periodic variability}
\label{sec:sec:per}

In \citetalias{2017Roquette}, we identified stars with periodic variability produced by the rotational modulation of a spotted stellar surface. As an initial step, the first morphological class defined was of candidate periodic stars. We considered a total of 1679 stars as candidate periodic because their light curves were visibly oscillating between max and min values, or because they presented significant peaks in their Lomb-Scargle periodogram \citep{lomb,scargle2}, in each photometric-band. In \citetalias{2017Roquette}, we measured periods for 1256 of these stars (75$\%$ of the candidate periodic sample), but due to varied contamination issues introducing ambiguities in the period determination process - such as limitations due to the time sampling - only 894 had reliable measured periods. Finally, periodicity seems more common among low amplitude variable stars, and only 31$\%$ of the candidate periodic stars were variables of type $S_1$. 

\begin{figure}
\centering
  \includegraphics[height=9cm]{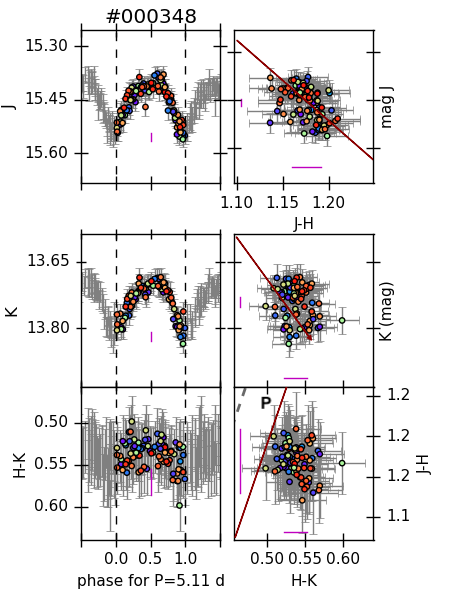}
  \caption{\label{fig:000348}Star CygOB2-000348 is an example of a diskless periodic variable, with S=1.62, a ptp amplitude of 0.13 and 0.11 mag in the J and K bands respectively, and that does not show periodicity in color. The plots, colors, and symbols are the same as in Fig. \ref{fig:093173}, but with the light curves and color curve phased with the period of the star, P=5.1d.}
\end{figure}

\begin{figure}
\centering
  \includegraphics[height=9cm]{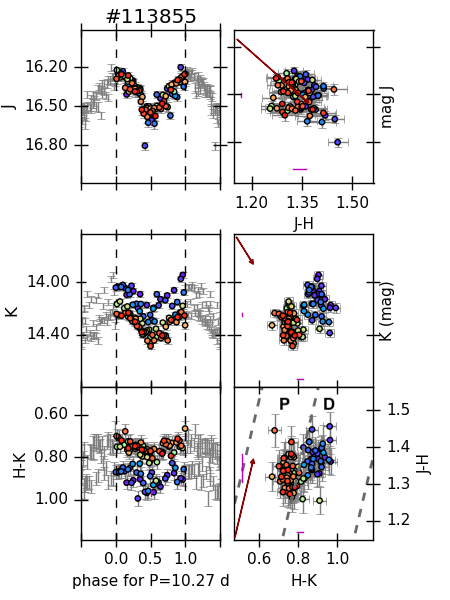}
  \caption{\label{fig:113855} Star CygOB2-113855 is an example of a disk-bearing (Class II) periodic star. With S=2.77 and ptp amplitude of 0.31 and 0.36 mag in the J and K bands, the star is also periodic in color. It is also an example of a compound variable with a variable IR-excess.  The plots, colors, and symbols are the same as in Fig. \ref{fig:093173}, but with the light curves and color-curve phased with the period of the star, P=10.3d.}
\end{figure}

Figs. \ref{fig:000348} and \ref{fig:113855} present two examples of periodic stars with their light curves folded to the periods obtained in \citetalias{2017Roquette}. The star 
CygOB2-000348 (Fig. \ref{fig:000348}) is diskless and periodic. The star CygOB2-113855 (Fig. \ref{fig:113855}) is disk-bearing and periodic in the light curves as well as in the color-curves. Only 39 stars ($\sim$2$\%$ of the candidate periodic sample) presented periodicity in their phase folded color curves. This lower incidence of periodicity in color does not mean that the physical mechanisms causing periodic variability are colorless, but their amplitudes of variability in the near-IR colors are often below our limits of detection.

\subsubsection{Occultation variability} 
\label{sec:sec:ec}

We based our definition of occultation variability on the existence of asymmetry in the light curve: these were stars that spent most of their observed time around the median value of magnitude but presented drops in brightness that can be periodic or not. Based on the visual inspection of the light curves, we identified 125 stars with 
occultation variability: 15 presented only a single or a couple of dips in their light curves; 110 presented multiple dips and we previously examined their periodicity in the context of \citetalias{2017Roquette}. 

\begin{figure}
  \includegraphics[height=9cm]{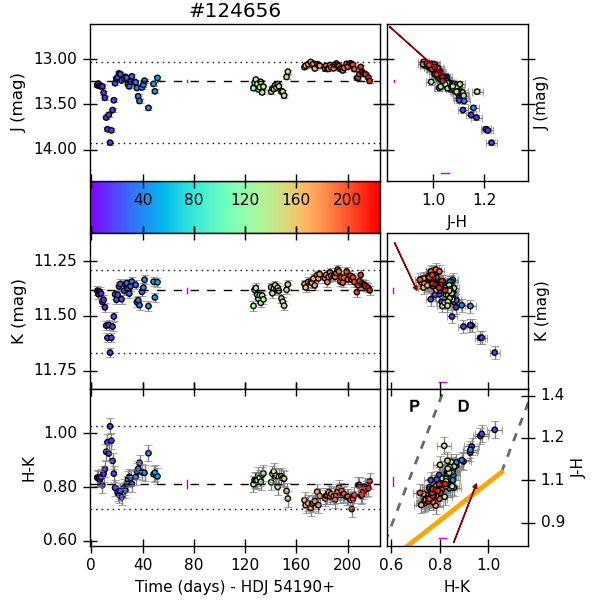}
  \caption{\label{fig:124656}Ddisk-bearing (Class II) star CygOB2-124656 is an example of an occultation variable that presents reddening linear trajectories in the color-color and color-magnitude diagrams. The star has S=3.59 and ptp amplitude of 0.43 and 0.17 in the J and K bands.  The plots, colors, and symbols are the same as in Fig. \ref{fig:093173}.}
\end{figure}

\begin{figure}
  \includegraphics[height=9cm]{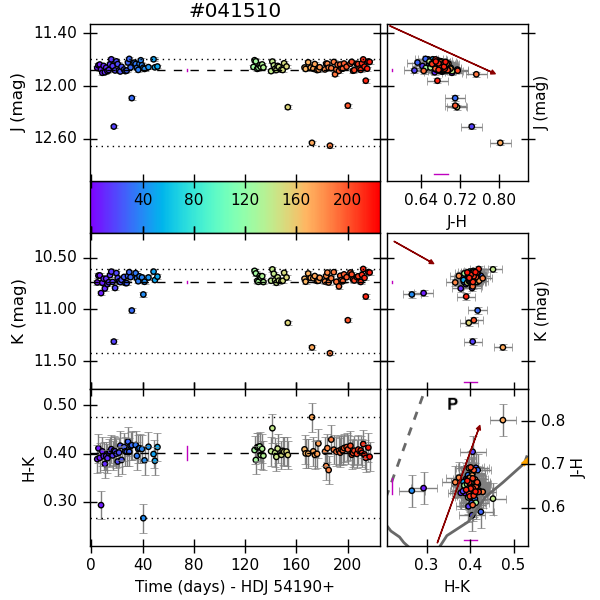}
  \caption{\label{fig:041510} Diskless star CygOB2-041510 is an example of a periodic occultation variable considered as an eclipsing binary candidate. It is also an example of a compound variable with occultation data-points with distinct color behavior. The star has $P=4.68$ days, $S=3.06$, and ptp amplitude of 0.28 in both J and K bands. The plots, colors, and symbols are the same as in Fig. \ref{fig:093173}.}
\end{figure}

Figs. \ref{fig:124656} and \ref{fig:041510} show two examples of occultation variables. The first is the disk-bearing star CygOB2-124656, which presented only two occultation features in the first season of observations. The second is the diskless star CygOB2-041510, which has periodic occultation features and is discussed in Sec. \ref{sec:sec:EB} as an eclipsing binary candidate.

 In \citetalias{2017Roquette}, we used Monte Carlo simulations to investigate the limitations of our data-sampling on detecting periodic stars with different waveforms. We concluded that our dataset was not appropriated for measuring reliable periods in periodic occultation wave-forms and that most of the periods measured were harmonics of the correct period. We, therefore, decided not to try to distinguish between eclipsing binary variables and other types of variables with occultation features (like dippers) based on their periodicity and level of asymmetry. Sec. \ref{sec:sec:EB} presents a further discussion about the identification of eclipsing binary candidates in our sample.

\subsubsection{Other types of variability}
\label{sec:sec:other}

\begin{figure}
  \includegraphics[height=9cm]{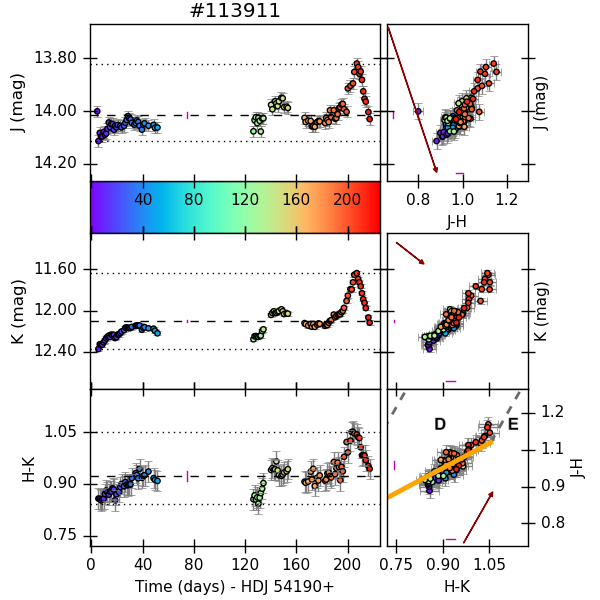}
  \caption{\label{fig:113911}Disk-bearing (flat-spectrum) star CygOB2-113911 is an example of a long-term variable with $S=3.67$, that presents a blueing linear trajectory in the color-magnitude diagrams. The plots, colors, and symbols are the same as in Fig. \ref{fig:093173}.}
\end{figure}

A total of 726 stars were identified as variables but presented neither periodicity nor
occultation features. Based on the visual inspection of the light curves, we verified the existence of 90 long-term variables (timescales longer than a month) as, for example, the star CygOB2-093173 in Fig. \ref{fig:093173}. We also identified 128 mixed-variables (variability with different time-scales) and 508 variables that were neither long nor mixed and were therefore called short-term variables (time-scales from days to weeks), for example, the star CygOB2-113911 in Fig. \ref{fig:113911}.

\subsubsection{light curve asymmetry}\label{sec:symetry}

We applied the asymmetry metric (M-index) as defined by \citet[][]{2014Cody} to statistically asses the level of asymmetry in the light curves. We defined a threshold for selecting asymmetric light curves by estimating the values of M-index in a large sample of sinewaves with the same time-sampling and uncertainty of our survey and verifying that all sinusoidal light curves were in the interval $-0.5\leq M\leq+0.5$.

We then used $M\!\geq\!+0.5$ to identify and flag 115 variables with dimming asymmetries. Of those, 67 were visually identified as occultation variables, including 13 of the 14 stars discussed as eclipsing binaries in Sec. \ref{sec:sec:EB}. Two examples are the stars CygOB2-124656 (Fig. \ref{fig:124656}) and CygOB2-041510 (Fig. \ref{fig:041510}). This criterion also flagged 29 periodic variables and 19 stars with other types of variability. We used $M\!\leq\!-0.5$ to select 11 stars with burst asymmetries. The stars CygOB2-093173 (Fig. \ref{fig:093173}) and  CygOB2-113911 (Fig. \ref{fig:113911}) are examples of it. All the stars with burst asymmetry were in the samples of other types of variability.

\subsection{Variability in color-space}

We mentioned in Sec. \ref{sec:sec:per} that around 2$\%$ of the candidate periodic sample also showed periodicity in their folded color curves. In this section, we identify other behaviors in the color space that could be used to infer the physical processes responsible for the variability. 

\subsubsection{Correlations between color and magnitude variations}
\label{sec:corrcor}

Variable stars may show strong correlations between their variability in different bands, which causes the star to move along linear trajectories in the color space. We investigated these correlations by measuring the slopes of the linear trajectories in the color-magnitude diagrams ($\frac{\Delta K}{\Delta (H-K)}$ and $\frac{\Delta J}{\Delta (J-H)}$) and in the color-color diagram ($\frac{\Delta (J-H)}{\Delta (H-K)}$).

As weak correlations may be biased by uncertainties in the data, before estimating the slopes, it was mandatory to  first investigate how strong was the correlation between the two variables in question. The Spearman nonparametric rank correlation test \citep{numericalr} was used to evaluate the statistical significance of the correlations in the color-color and color-magnitude diagrams. Spearman correlation measures how well the relationship between two variables can be described by a monotonic function. It can assume values between -1 and 1 with the extremes occurring when the relationship between the variables is perfectly monotonic. For stars with the absolute value of Spearman correlation larger than 0.7, the orthogonal distance regression method\footnote{Both Spearman nonparametric rank correlation and Orthogonal Distance Regression are included in the SciPy package for Python 2.7} was used to estimate the slopes of the trajectories in the color-color and color-magnitude diagrams and their uncertainties. This method takes the uncertainty of each data point into account while performing a linear fit. Finally, we overplotted the slopes measured along with the real data in each of the color spaces, and visually inspected them in order to remove stars that were not well described by a linear relation.  

\begin{figure*}
\centering  \includegraphics[width=15cm]{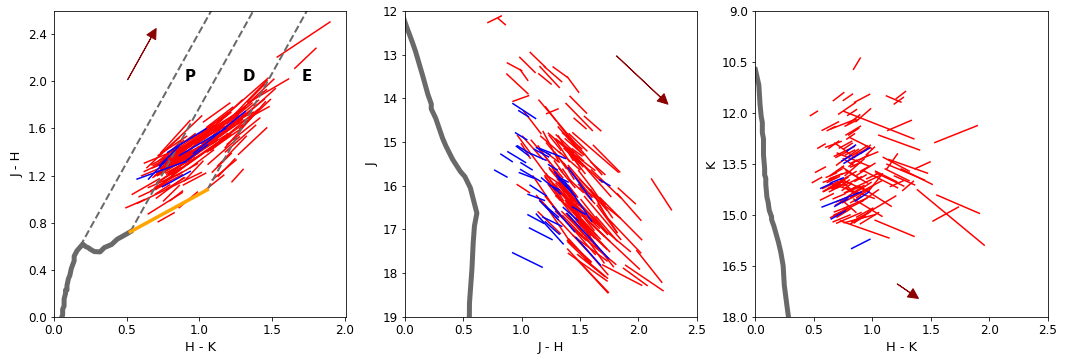}
  \caption{\label{fig:CCslopes} Color-magnitude and color-color diagrams with the slopes for the stars presenting a linear variation. Red lines show disk-bearing stars, and blue lines show stars without discs. The length of each line segment reflects the amplitude of the variability. The dark red arrows show the median interstellar extinction toward CygOB2, and the gray lines show as reference of unreddened normal photosphere colors, an empiric isochrone from \citet{2007Kraus}.}
\end{figure*}

Fig. \ref{fig:CCslopes} and Table \ref{tab:slopes} present the slopes for the 335 variable stars with trajectories in color space that could be well fitted by a linear relation. The majority of those stars showed higher ptp amplitudes than the median values for the whole sample of variable stars, or belonged to the $S_1$ sample. This suggests that our analysis of the correlations between color and magnitude variability is biased toward high amplitude variable stars.

Only 50, out of 335 stars with linear trajectories were diskless. Among the 144 stars with measured $\frac{\Delta K}{\Delta (H-K)}$ slopes, 29 had positive slopes and 115 had negative slopes. Among the 196 with measured $\frac{\Delta J}{\Delta (J-H)}$ slopes, 193 had positive slopes, and 3 had negative slopes. Finally, 143 stars had measured $\frac{\Delta J-H}{\Delta (H-K)}$ slopes and all of them positive.

\begin{table}
  \addtolength{\tabcolsep}{-1.0pt}
      \caption{\label{tab:slopes}Parameters for the stars with measured slopes in the color-color diagram or color-magnitude diagrams. The full table is available online at the CDS.}
      \centering
      \begin{tabular}{p{1.2cm}|p{1.0cm}p{0.8cm}p{1.0cm}p{0.79cm}p{0.99cm}p{0.69cm}}
        \hline
        \hline
         &  \multicolumn{6}{|c}{Slopes ($^\mathrm{o}$)} \\
        
        \small{ID} &  \tiny{$\frac{\Delta J-H}{\Delta (H-K)}$}&  \small{err} &
        \tiny{$\frac{\Delta K}{\Delta (H-K)}$} &  \small{err} &  \tiny{ $\frac{\Delta J}{\Delta (J-H)}$} & \small{err}\\        
        \hline
  000001  &      &         &       &       &51.7   &  5.7    \\  
  000012  &      &         &       &       &66.1   &  2.2    \\  
  000047  &      &         & -69.5 &2.0    &       &         \\  
  000064  & 42.6 &    0.2  &       &       & 73.1  &  0.9    \\  
        \hline
      \end{tabular}
\end{table}

The stars CygOB2-124656 (Fig. \ref{fig:124656}), CygOB2-093173 (Fig. \ref{fig:093173}) and CygOB2-113911 (Fig. \ref{fig:113911}) are examples of 
stars presenting linear trajectories in the color-diagram. We further discuss in Sec. \ref{sec:var} how the slopes of their trajectories can help distinguishing between the different possible physical mechanisms causing variability.

In the discussion that follows, it is crucial to understand how the slopes are defined. In the case of color-magnitude diagrams (Left and middle plots in Fig. \ref{fig:CCslopes}), a positive slope is defined as a positive change in color, and a positive change in magnitude, which results in a vector pointed toward the bottom-right of the color-magnitude diagram due to the inverted scale of magnitude, and the slope grows clockwise. In the case of the color-color diagram (Right plot in Fig. \ref{fig:CCslopes}), a positive slope is defined as positive changes in both colors, the resultant vector points to the top right, and the slope grows counterclockwise.

\subsubsection{Compound variables}
\label{sec:sec:comp}

Often the variability behavior was not confined into a single monotonic trajectory with a positive or negative slope in the color-space. For example, the disk-bearing periodic star CygOB2-113855 (Fig. \ref{fig:113855}) has an approximately fixed amplitude in its periodic variability, but presents a decreasing flux in K responsible for a varying near-IR excess that shifted the star across the K vs. H-K and color-color diagrams. Such a complex color behavior may indicate multiple variability mechanism in action in the same star. \citetalias{2015Rice} describes similar behaviors as a ``compound color behavior''. We verified three types of compound color behavior in our sample:

\emph{Asymmetry with a distinct color:} The 
occultation variable CygOB2-041510 (Fig. \ref{fig:041510}) presents occultation datapoints with distinct color behavior than the rest of the stellar points. Similarly, 20$\%$ (25) of the occultation variables presents similar behavior, which is associated with eclipsing binary variability in Section \ref{sec:sec:EB}. 

\begin{figure}
  \includegraphics[height=9cm]{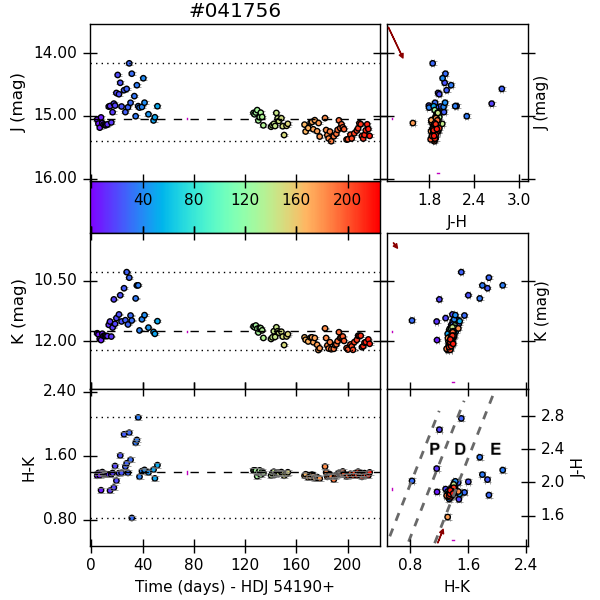}
  \caption{\label{fig:041756}Disk-bearing (Class II) star CygOB2-041756 (S=10.4) is an example of a mixed variable with a long term variability mixed with burst features in the first epochs observed. The star is a compound variable with asymmetries in the light curve having a distinct color behavior. The plots, colors, and symbols are the same as in Fig. \ref{fig:093173}.}
\end{figure}

We also verified this behavior in variable stars with burst asymmetry in the light curve. For example, the star CygOB2-041756 (Fig. \ref{fig:041756}) shows a mixture of variability modes. Overall, the star is dimming while describing a linear trajectory with a negative slope in the CMD. However, in the first 50 epochs of observation, the star has several burst-features with varying amplitudes up to one magnitude in both J and K, showing a distinct behavior in the color space.

\emph{Variable near-IR excess:} some stars present an increasing or decreasing near-IR excess, causing the data points observed in different seasons to be clumped in different regions of the color-color and CMDs. The periodic star CygOB2-113855 (Fig. \ref{fig:113855}) is an example of it: while the amplitudes of periodic variability are unchanged during the observed time, and no significant change is observed in the J band, the median magnitude in the K band diminishes shifting from the ``P'' to the ``D'' region of the color-color diagram. We note that whatever is the physical mechanism causing the decrease of brightness in the K band, it seems to be independent of the dominant physical mechanism producing the periodic variability. This type of variability can be associated with what \citet{2012Rice} called ``transient near-IR excesses'', which is related to a variability mechanism that changes the amount of near-IR excess and some times makes the star transition between the ``P'', ``D'' and ``E''  regions of the color-color diagram.

\begin{figure}
  \includegraphics[height=9cm]{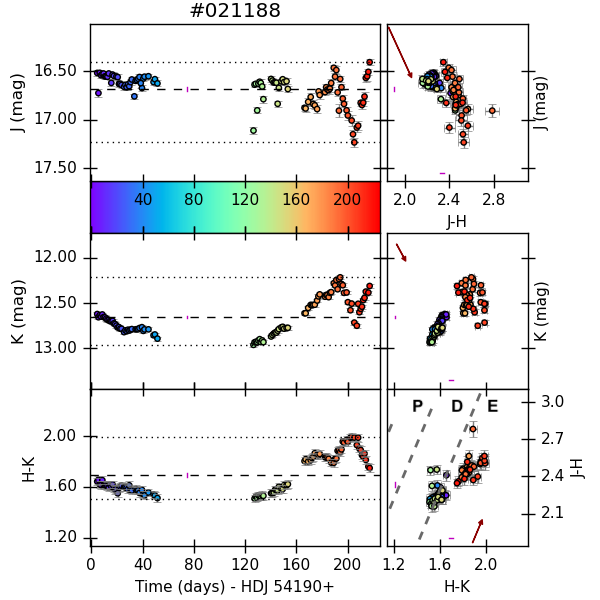}
  \caption{\label{fig:021188}Disk-bearing (Class II) star CygOB2-021188 (S=1.0) is an example of compound variability showing a variable near-IR excess and a changing slope during the observed time. The plots, colors, and symbols are the same as in Fig. \ref{fig:093173}.}
\end{figure}

\emph{Changing slope:}
some stars show an evident change in the dominant variability mechanism, marked by a change in the slope of variability in the color-diagram. For example, the star CygOB2-021188 (Fig. \ref{fig:021188}) gets bluer when fainter during the first $\sim$100 observed nights, and it presents a linear trajectory with a negative slope in the K vs. H-K CMD during this time. However, during the final $\sim$100 nights of observations, the star presents a shorter term variability in which it gets redder when fainter, and it presents a linear trajectory with a positive slope in the J vs. J-H CMD during this time. Note that the star also shows a variable near-IR excess. As a consequence, it transitions between the ``D'' and ``E'' parts of the color-color diagram during the observed time.

\begin{figure*}
    \includegraphics[width=17cm]{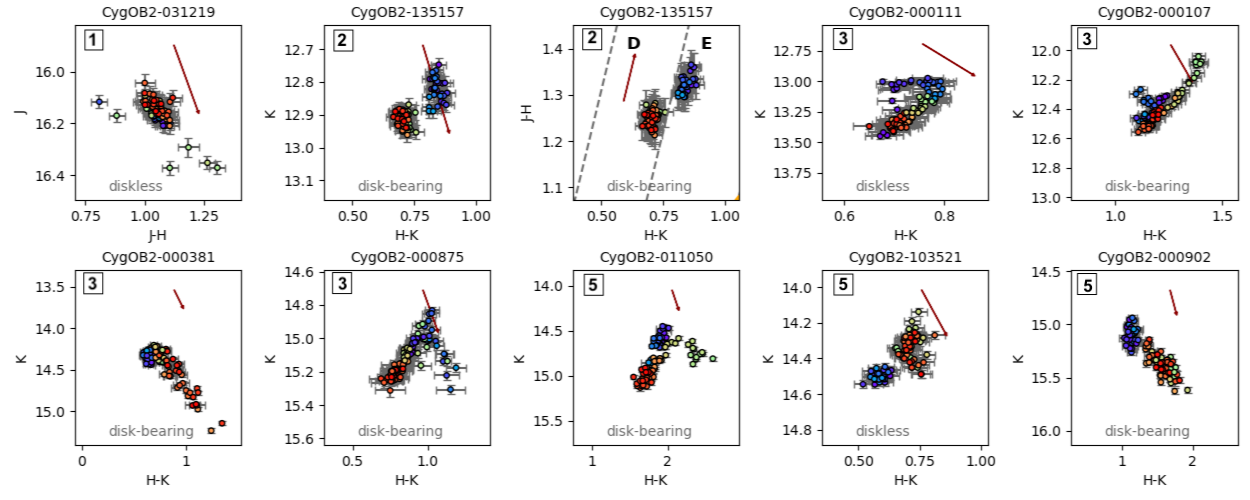}
  \caption{\label{fig:compound}Examples of stars with a compound behavior are shown inside color-color diagrams or color-magnitude diagrams. 1- Asymmetry with a distinct color; 2- Variable IR-excess; 3- Changing slope; 5- Mix of 2 and 3. 
  Dark red arrows show the effect of interstellar reddening for $A_V=1$ mag. The color of each data point reflects its observational date following the color-bar in Fig. \ref{fig:093173}.}
\end{figure*}

We identified 149 stars with one or multiple modes of compound variability. Fig. \ref{fig:compound} shows more examples of each case. As in star CygOB2-021188 (Fig. \ref{fig:021188}), often variable near-IR excess and changing slope happened at the same time. 

\section{Results and discussion}
\label{sec:discussion}

The results obtained for the 2529 variable stars studied in this paper are available as electronic tables at CDS. Table \ref{tab:data} presented general information on the light curves used in this study, along with literature comparison for each variable star on our list. Table \ref{tab:data2} presented the information related to the variability classification. Table \ref{tab:resumo} summarizes how we categorized disk-bearing and diskless stars as discussed in the previous sections.

\begin{table*}
  \centering
  \addtolength{\tabcolsep}{-1.0pt}  
  \caption{\label{tab:resumo}Summary of the variability classification performed in Sec. \ref{sec:results} for both disk-bearing and diskless samples. The amount of stars falling in each class is shown along with the number of stars in $S_1$ and $S_2$ samples, the number of rejected $S_2$ variables after the $\chi^2$-test and the number of stars with measured slopes in each diagram.}
  \footnotesize{
    \begin{tabular}{p{1.7cm}p{0.6cm}p{0.5cm}p{0.5cm}|p{0.37cm}p{0.4cm}p{0.37cm}p{0.42cm}p{0.42cm}p{0.42cm}p{0.4cm}p{0.37cm}p{0.35cm}p{0.5cm}|p{0.7cm}p{0.7cm}p{0.8cm}|p{0.65cm}p{0.65cm}}
\hline
\hline
      & Total & Non & Rej & \multicolumn{2}{c}{eclipse} & \multicolumn{2}{c}{periodic} &
      \multicolumn{6}{c}{Other types} &
     \multicolumn{3}{|c}{Measured Slopes} &
      \multicolumn{2}{|c}{Total var}  \\ 
      & & Var& & \multicolumn{2}{c}{like}  & 
      \multicolumn{2}{c}{candidates} & 
      \multicolumn{2}{c}{ShortTerm} &  
      \multicolumn{2}{c}{Long Term} &
      \multicolumn{2}{c|}{Mixed} &
      \multicolumn{3}{c|}{} & 
      \multicolumn{2}{c}{}  \\ 
      &  & & & $S_1$ & $S_2$   & $S_1$ & $S_2$  & $S_1$ & $S_2$  & $S_1$
      & $S_2$  & $S_1$ & $S_2$  & \tiny{$\frac{\Delta J-H}{\Delta (H-K)}$} & \tiny{$\frac{\Delta K}{\Delta (H-K)}$} &  \tiny{$\frac{\Delta J}{\Delta (J-H)}$}  &   $S_1$ & $S_2$\\
    \hline
      Disk-bearing  & 1272 & 180 & 272 & 89 & 0  &133 & 38     & 328 & 33 & 72 & 14 & 104 & 9 & 136 & 134 & 157  &
      726 & 94 \\
      Diskless   & 3811 & 824 &1278 & 30  & 5  &392 & 1116   & 114 & 33 & 4  & 0  & 14 & 1  & 8 & 10 & 56  &
      554 & 1155\\
        \hline
\end{tabular}}
\end{table*}

\subsection{Variability trajectories in the color-space and the physics behind photometric variability of young stars}\label{sec:var}

\begin{figure}[htb]
 \includegraphics[width=9cm]{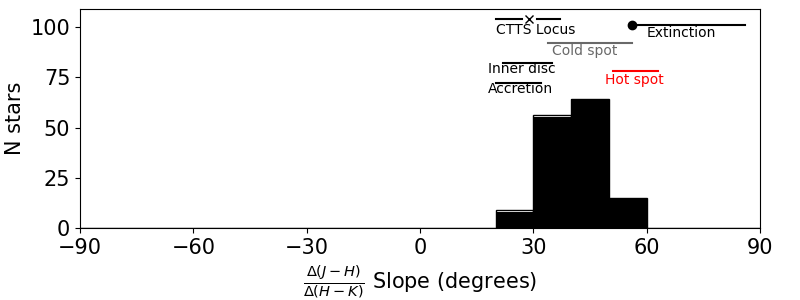}\\
 \includegraphics[width=9cm]{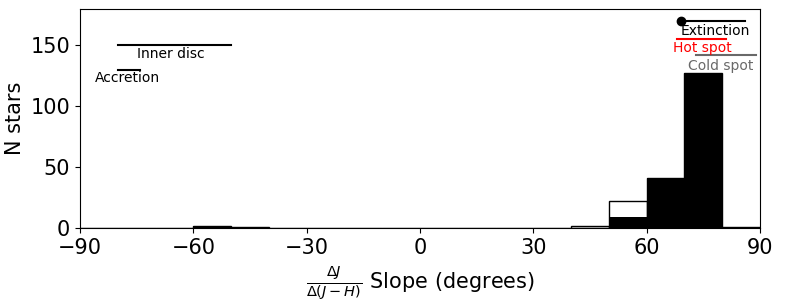}\\
   \includegraphics[width=9cm]{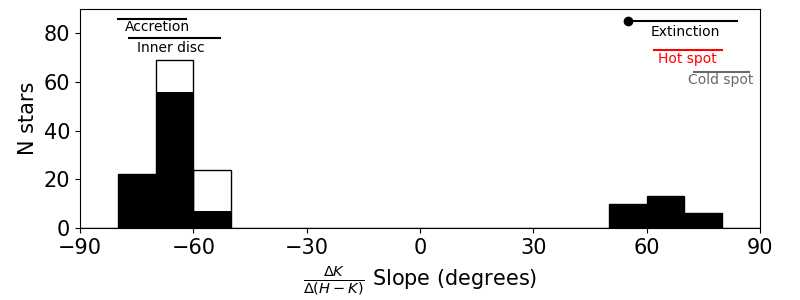} 
   \caption{\label{fig:slopes}Distribution of slopes measured in the color-color and color-magnitude diagrams. The figure shows slopes for J-H vs. H-K (top), J vs. J-H (middle), and K vs. H-K (bottom) diagrams. We plotted the expected slopes for each physical mechanism as line segments, following the ranges in Table \ref{tab:slopetheory}. The X mark shows the exact position of the T Tauri Locus. The black circle shows the slope expected for the typical dust opacity index in the ISM. Black bars show the stars in $S_1$ sample and with precision in the light curve better than 5$\%$, empty bars show stars in the $S_2$ sample or with precision worse than 5$\%$.}
\end{figure}

In Sec. \ref{sec:corrcor} we estimated the slopes of the linear trajectories described by 355 variable stars inside the color diagrams in Fig. \ref{fig:CCslopes}. In this section, we put such slopes in context by comparing them to reference values collected in the literature and discussing their relation with the physical mechanisms responsible for variability in the near-IR. Fig. \ref{fig:slopes} shows the distributions of  $\frac{\Delta J-H}{\Delta (H-K)}$ , $\frac{\Delta J}{\Delta (J-H)}$, and $\frac{\Delta K}{\Delta (H-K)}$ slopes measured in Sec. \ref{sec:corrcor}. When examining Fig. \ref{fig:slopes}, we stress that the distributions of slopes in each color space do not necessarily correspond to the same stars. Out of 355 stars that had at least one slope measured: 20 stars had all three slopes measured; 12 had both $\frac{\Delta J-H}{\Delta (H-K)}$ and $\frac{\Delta K}{\Delta (H-K)}$, but no $\frac{\Delta J}{\Delta (J-H)}$; 88 had both $\frac{\Delta J-H}{\Delta (H-K)}$ and $\frac{\Delta J}{\Delta (J-H)}$, but no $\frac{\Delta K}{\Delta (H-K)}$; nine had both  $\frac{\Delta K}{\Delta (H-K)}$ and $\frac{\Delta J}{\Delta (J-H)}$, but no $\frac{\Delta J-H}{\Delta (H-K)}$; 24 had only $\frac{\Delta J-H}{\Delta (H-K)}$; 79 had only $\frac{\Delta J}{\Delta (J-H)}$; 103 had only $\frac{\Delta K}{\Delta (H-K)}$.

\begin{table}[htb]
  \centering
  \addtolength{\tabcolsep}{-1.2pt}
  \caption{\label{tab:slopetheory}Slopes of variability in the near-IR expected from models and other observational studies.}
  \begin{tabular}{p{2.05cm}p{1.5cm}p{1.4cm}p{1.4cm}p{.9cm}}
    \hline
    \hline
    Physical &  $\frac{\Delta K}{\Delta (H - K)}$ & $\frac{\Delta
    J}{\Delta (J - H)}$ &  $\frac{\Delta (J - H)}{\Delta (H - K)}$ &
    Ref\footnotemark[1] \\
 Mechanism & $\;\;\;$($^\mathrm{o}$) &  $\;\;\;$($^\mathrm{o}$) & $\;\;\;$($^\mathrm{o}$) & \\
\hline
    hot spot\footnotemark[2]                  & $\;$63 : $\;$79 & $\;$69 : $\;$80 & 52 : 62 & 1,2,3 \\
    cool spot\footnotemark[2]                  & $\;$74 : $\;$86 & $\;$74 : $\;$88 & 35 : 55 & 1,2  \\
    extinction\footnotemark[2]                 & $\;$58 : $\;$83 & $\;$70 : $\;$85 & 58 : 85 & 1,2 \\
    variable accretion rate   & -54 : -76 & -51 : -79 & 21 : 31 &  2  \\
    Changes in the inner disk & -63 : -79 & -75 : -79 & 23 : 34 & 2 \\
    \hline
  \end{tabular}

  \footnotemark[1]{Ref: Reference for the values presented are 1, for this study; 2 for \citetalias{2001Carpenter}; 3 for \citet{2009Scholz}.}
  \footnotemark[2]{A description of our estimations are presented in Appendix \ref{app:A}}
\end{table}

\subsubsection{Positive slopes in the CMD: spot and extinction variability}

Stellar spots and extinction variability can cause positive slopes in the near-IR CMD. Stellar spots can be hotter or colder than the stellar photosphere. They produce variability because the fraction of the stellar surface covered by the spots and seen by the observer changes as the star rotates. Cold spots are similar to sunspots \citep[e.g.,][]{1993Bouvier}, they are colder than the photosphere and arise from magnetic activity. In young stars, they can cover up to $\sim$60$\%$ of the stellar surface \citep{2015Venuti}. They can be very stable and persist over thousands of rotation periods \citep{Herbst1994}. Cold spots are likely the most common source of photometric variability in low-mass stars at all ages. Indeed, in \citetalias{2017Roquette} we found 1119 diskless stars with periodic variability likely associated to cold spots, which accounts for 65$\%$ of the diskless variables in our study.

Hot spots are only possible in disk-bearing accreting stars as they form in the regions where the accretion streams hit the stellar surface. Stable accretion streams are generated by a misalignment between the rotation and magnetic axis, and the hot spots are typically located at high latitude \citep{2008Kulkarni,2013Romanova}. Unlike cold spots, they only persist over a few rotation periods. In case of unstable accretion, hot spots may be located at low latitudes, and the accretion bursts occur stochastically and last only a few hours, typically less than a stellar rotational period \citep{2012Romanova,2013Kurosawa,2014Stauffer,2016Colombo}.

With cold spots, the star gets redder and fainter when the spot is visible. With hot spots, the star gets bluer and brighter when the spots are visible. In both cases, the resultant trajectory in the CMD will have a positive slope. The amplitude and slope of the trajectories in the color-space depend on the difference of temperature between the spot and the star, and also on the size of the spot, but cold and hot spots are expected to produce a similar range of slopes in the CMDs. On the color-color diagrams, cold spots usually describe shallower trajectories than hot spots.

In Appendix \ref{app:A:spot},we used a spot model to infer typical near-IR amplitudes and slopes for spot variability. We fed the model with values of spot temperatures in YSOs inferred from optical and UV observations in the similarly aged cluster NGC 2264 \citep{2015Venuti}, along with the effective temperature of the same stars, and obtained expected amplitudes in the JHK bands. We present the range of slopes found in Table \ref{tab:slopetheory}, along with literature values.

Extinction variability can arise from inhomogeneities in the absorbing material moving across the line of sight. These inhomogeneities can be in the inner circumstellar environment or in the ambient molecular cloud. For example, in AA Tauri variable stars \citep[e.g.,][]{2007Bouvier}, the magnetospheric star-disk interaction, via which accretion occurs, can lift some dust above the plane of the disk, creating a warp in the inner disk. As the dust is optically thick, this warp causes extinction occultation of the stellar photosphere.

As with spots, extinction variability also makes the star redder as it gets fainter. The slopes of the trajectories described by the cold spots, hot spots and extinction can be very similar, making it difficult to distinguish between the three phenomena in the color-space. Our estimations of typical extinction slopes are presented in Table \ref{tab:slopetheory} and detailed in Appendix \ref{app:A:ext}. In the remaining discussion, we call stars describing linear trajectories with position slopes ``reddening variables''.

For example, the disk-bearing star CygOB2-093173 (Fig. \ref{fig:093173}) presents a reddening variability with a 42.9$^\mathrm{o}$ slope in the J-H vs. H-K diagram, 74$^\mathrm{o}$ in the J vs. J-H CMD and 55$^\mathrm{o}$ in the K vs. H-K CMD. The star has high ptp amplitudes (1.79 in J and 0.77 in K), and has a borderline compound behavior in the K vs. H-K diagram, where the positive slope of its trajectory became shallower after the 80 first nights of observation, suggesting a mix of spot variability and extinction.

The disk-bearing occultation variable star CygOB2-124656 (Fig. \ref{fig:124656}) is another example of reddening variability. Over the course of the occultation feature (first $\sim$80 nights of observation), the star presents a reddening slope, which is more remarkable in the J vs. J-H diagram, with a 72$^\mathrm{o}$ inclination. In the K vs. H-K diagram, the star does not present a definite linear trajectory in the second season of the observations when occultation features are no longer observed. The reddening variability correlated to the occultation features, added to the fact the star has a disk, suggest the variability is caused by dust and the star is likely an AA Tauri variable.

\subsubsection{Negative slopes in the CMD: disk emission and accretion variability}
\label{sec:dema}

Disks can contribute to the IR emission of YSOs by absorbing the optical and ultraviolet radiation from the star and reradiating it at longer wavelengths, and by releasing energy as the material in the disk is radially transported through a viscous optically thick accretion disk \citep{1974Lynden-Bell,1997Meyer}. The resultant IR-excess can occur in different wavelengths from near-IR to far-IR. The region in the star-disk system producing an excess in each wavelength range depends on the temperature distribution in the star-disk system. Typically, the near-IR excess originates in the hotter inner parts of the disk, and the shorter wavelength at which the IR-excess is observed depends both on the temperature of the inner disk material and on the inner hole size. An increasing accretion rate, for example, changes the release of energy and temperature distribution in the inner disk which alters the amount of absorbed and reprocessed radiation \citep{2009Scholz}. As the temperatures in the inner disk become larger than the stellar surface temperature, the flux from the disk overcomes the stellar one and the star gets brighter in the near-IR and redder. On the other hand, holes in the inner disk alters the amount of absorbed and reprocessed radiation, diminishing the disk contribution to the overall near-IR flux. As the disk emission becomes stronger at longer wavelengths, one can expect higher amplitude variations in the K band, than in the J band, while the variations make the star bluer (approaching the true stellar color) as it fades.

In both cases, variability related to the disk emission will produce negative (``blueing'') slopes inside the color-magnitude diagrams and  positive slopes inside the color-color diagram. Table \ref{tab:slopetheory} shows literature values for the slopes produced by these type of physical mechanisms. \citetalias{2001Carpenter} reports amplitudes as large as 1 magnitude in K band for this type of variability.

The disk-bearing star CygOB2-113911 (Fig. \ref{fig:113911}) is an example of blueing variable. The star described linear trajectories in all diagrams with a 48.72$^\mathrm{o}$ slope in the J-H vs. H-K diagram, -74$^\mathrm{o}$ in the K vs. H-K diagram, and -49$^\mathrm{o}$ in the J vs. J-H diagram. The datapoints dominating the linear trajectory are related mainly to two burst-features in the light curve: a lower amplitude one between nights 120 and 160, and a very intense one after night 170. The behavior of these burst features suggests that its variability is caused by variable accretion episodes.

\subsubsection{Distributions of slopes}

Fig. \ref{fig:slopes} showed the distribution of slopes of the linear trajectories described by 355 variable stars inside the color diagrams. The distributions can be compared to the ones for the ONC in \citetalias{2001Carpenter} (see their Fig. 20) and \citetalias{2015Rice} (see their Fig. 20). To facilitate the comparison, we adopted the same terms as these authors, and directly compare reddening variables (positive slopes in the CMDs related to spot or extinction variability) with blueing variables (negative slopes in the CMDs related to variability caused by changes in the inner disk or variable accretion).

In the distribution of slopes in  the J-H vs. H-K diagram (top panel in Fig. \ref{fig:slopes}), the distinction between types of variability is more complicated, as all physical mechanisms investigated produce positive slopes inside the color-color diagram. Even though blueing variability is expected to produce slopes shallower than about 35$^\mathrm{o}$, there is still a significant overlap between the slopes of spotted stars and the TTauri locus. With 28 stars with shallower slopes in the range 21$^\mathrm{o}$-35$^\mathrm{o}$ and 116 in the range 35$^\mathrm{o}$ and 58$^\mathrm{o}$, this diagram indicates a mixture of disk blueing variability and spot variability.

In the J vs. J-H diagram (middle panel of Fig. \ref{fig:slopes}), we found mostly reddening stars, with 154 stars presenting variability with slopes between 67$^\mathrm{o}$ and 81$^\mathrm{o}$, and only three stars presenting blueing variability with slopes between -49$^\mathrm{o}$ and -58$^\mathrm{o}$. This indicates that this diagram is composed of a mixture of extinction and spot variability. We found 39 stars with slopes shallower than 67$^o$, which is a larger number of shallow positive slopes than \citetalias{2001Carpenter} or \citetalias{2015Rice}. However, most of those are from the $S_2$ sample or had large photometric uncertainty between 5 and 10$\%$.

In the K vs. H-K diagram (bottom panel of Fig. \ref{fig:slopes}), we found four times more blueing variables than reddening ones, with 115 stars presenting blueing variability with slopes between -51$^\mathrm{o}$ and -80$^\mathrm{o}$, and 29 stars presenting reddening variability with slopes between 55$^\mathrm{o}$ and 74$^\mathrm{o}$. This distribution differs a lot from \citetalias{2015Rice}, where they found only 1.2 times more blueing variables than reddening ones. As \citetalias{2015Rice} used $S\!\geq\!1$ for selecting variable stars, if we considered in the comparison only $S_1$ sources with photometric uncertainty lower than 5$\%$ (black bars in Fig. \ref{fig:slopes}), we still find three times more blueing variables than reddening variables.

The fact that few stars presented blueing variability in the J vs. J-H CMD is explained by the nature of the near-IR excesses in disk-bearing stars. As the amount of light re-emitted by the disk in the near-IR is larger for longer wavelengths, one may expect that the amplitudes of variability due to variable accretion rate and changes in the inner disk will also be more significant for the K magnitude and H-K color, than for J magnitude and J-H color, thus also explaining why more blueing variability is observed in the K vs. H-K diagram. The flux and variability in the J band is dominated mainly by the stellar flux, therefore explaining why the distribution of slopes in the J vs. J-H CMD is dominated by reddening related to spot variability, which acts directly in the photosphere, and extinction, which dims the light emitted by the star itself.

Considering both J vs. J-H and K vs. H-K diagrams, \citetalias{2001Carpenter} found six times more reddening type variables than blueing type variables, \citetalias{2015Rice} found 1.5 times more reddening than blueing types. Considering only stars in the $S_1$ sample with photometric uncertainty lower than 5$\%$, our dataset presents more similar results to \citetalias{2015Rice} than with \citetalias{2001Carpenter}, as we found 2.1 times more reddening than blueing variables.
We caution the reader on interpreting the slope distributions. While the study of the trajectories in color is especially useful for selecting high amplitude variable stars with a single dominant physical mechanism producing variability in their light curves, the frequency in which each physical mechanism is observed in a dataset is most likely not representative of the recurrence of such physical mechanism in nature. \citetalias{2015Rice} remarked that the larger fraction of blueing variables that they found compared to \citet{2001Carpenter} may be because the differences in the total time-span of each survey may favor the detection of certain types of variables. For example, surveys developed over a couple of months, like \citet{2001Carpenter} and \citet[][who surveyed NGC 2264 for variability in the optical and mid-IR]{2014Cody} might favor the detection of variable stars such as AA Tauri, which present variability with time-scales of days to weeks. On the other hand, as blueing variability has typically long time-scales, surveys as long as \citet[][2.5 years]{2015Rice}, \citet[][1.5 years]{2012Rice} or the present study may favor the detection of this type of variable.

We identified four major differences between our data analysis and other studies that could introduce bias to our results and explain the differences with results from the literature. First, in our survey we considered stars with photometric uncertainties up to 10$\%$. Second, even when considering only stars with uncertainty lower than 5$\%$, our survey goes deeper than \citetalias{2015Rice} by 0.2 mag in the J band, and by 0.3 mag in the H and K bands. Third, a collateral effect of the definition of our $S_2$ sample is the introduction of variables with large amplitude, but with high uncertainties. Fourth, both \citetalias{2015Rice} and \citetalias{2001Carpenter} searched for variability among $\sim$18000 good quality light curves in the FOV of ONC. In our case, we only analyzed light curves extracted for previously known YSO in CygOB2. As discussed in Sec. \ref{sec:dema}, the variability mechanisms producing blueing variability are intimately related to the infrared-excess in disk-bearing stars. This means that by analyzing the light curves for disk-bearing stars identified by GDW13 we are preferentially selecting stars that have augmented IR-excess and more likely to show blueing variability.

We address the first and third points by looking only at the stars in the $S_1$ sample with uncertainty lower than 5$\%$, which accounts for 26 reddening stars and 84 blueing stars in the K vs. H-K diagram, still yielding 3.2 times more blueing variables. Therefore confirming our previous results. On the other hand, the second and fourth points are likely the most important ones to keep in mind when comparing our results with the literature, specially when looking at the K vs H-K CMD, which is mostly dominated by disk-variability. \citet{2015Rice} found 1.2 times more blueing variables in this diagram than reddening ones, in a sample of $\sim$15k stars in the FOV of ONC. We found 3 more times blueing variables in a sample of $\sim$5k YSOs, 1/4 of which known to have strong IR-excesses.

\subsection{Eclipsing binary candidates}\label{sec:sec:EB}

Another source of variability not yet discussed is the occurrence of eclipses in a binary system. We verified the occurrence of eclipsing binaries in our dataset in two ways:
First, we further analyzed our list of occultation variables to identify eclipsing binary candidates based on their variability characteristics. Second, we matched our dataset with the list of eclipsing binaries in the field of view of CygOB2 previously identified by \citetalias{hend2011+}. 

We considered that an occultation feature was due to an eclipsing binary if:
\begin{enumerate}
\item The star was classified as occultation variability (Sec. \ref{sec:sec:ec});
\item The drops in brightness were sharp, deep and homogeneous:
\begin{enumerate}
\item Given the combination of the cadence of about one observation per night in each filter, and the short periods expected for eclipsing binaries\footnote{The detectability of a binary system as an eclipsing binary depends on the inclination of the system and its orbital period and separation. Thus the chances of detecting eclipses are higher for shorter orbital periods, because the wider the separation of components is, the longer the orbital period gets, and the narrower is the range of inclinations in which the eclipse is visible. This way, we expect most of the eclipsing binaries detected to have short periods.}, each drop in brightness is expected to be composed typically of one or two sets of JHK observations;
\item The eclipse happened similarly in the three bands;
\item Their typical depth was $\gtrsim$0.1 mag ;
\end{enumerate}
\item The star was classified as a compound variable of the first type (Sec. \ref{sec:sec:comp}), with all the data points that are part of the eclipse being localized outside the major bulk or pattern of observations inside the CMDs, and not following a correlated slope as in Sec. \ref{sec:corrcor}. 
\end{enumerate}

\begin{table*}[htb]
    \centering
    \begin{tabular}{ccccccccccc}
    \hline\hline
 ID  & HSP & J & H & K & ptpJ  &ptpH&  ptpK & Type & Per & Per$_\mathrm{rot}$ \\
  (CygOB2-) & & (mag) & (mag) & (mag) & (mag) & (mag) & (mag) & & (d) & (d) \\
\hline
  000644&   21 &  13.68 & 13.14 & 12.93 & 0.08 & 0.08 &  0.08 &  3  & 2.89 (1.44/3.21) &  - \\ 
  010997&      &  11.54 &  10.88  &  10.48 & 0.10 & 0.11 &  0.10 &  3  &  -  &  -  \\ 
  041491&      &  14.31 &  13.23  &  12.73 & 0.16 & 0.16 &  0.17 &  3  &  1.58/3.77 & 7.39  \\ 
  041510&      &  11.80 &  11.10  &  10.70 & 0.28 & 0.23 &  0.28 &  3  &  4.69 & -  \\ 
  041524&      &  16.29 &  15.03  &  14.45 & 0.17 & 0.16 &  0.19 &  3  &   - & -  \\ 
  041588&      &  10.45 &   9.97  &   9.66 & 0.25 & 0.23 &  0.21 &  3  &  1.11 & -  \\ 
  072104&  58  &  14.59 &  13.95  &  13.65 & 0.33 & 0.30 &  0.31 & 3   &  0.83 & -  \\
  072174&  106 &  15.28 &  14.22  &  13.74 & 0.21 & 0.23 &  0.25 &  3  &  7.30 & -  \\ 
  103439&      &  11.64 &  11.24  &  10.98 & 0.14 & 0.18 &  0.15 & 3   &  - &  - \\
  103616&      &  15.52 &  14.44  &  13.93 & 0.26 & 0.23 &  0.22 & 3   &  1.14 & -  \\
  114036&      &  15.54 &  14.28  &  13.67 & 0.16 & 0.14 &  0.15 & 3   &  0.90 &  - \\
  114133&      &  16.18 &  15.06  &  14.54 & 0.33 & 0.27 &  0.30 &  3  &  1.97/3.95 &   \\ 
  124487&      &  14.51 &  13.40  &  12.88 & 0.22 & 0.21 &  0.20 &  3  & 4.77  &  1.12 \\ 
  134771&      &  12.80 &  12.04  &  11.61 & 0.06 & 0.06 &  0.07 &  3  & -  & -  \\ 
  114152&  92  &  15.34 &  14.54  &  14.12 & 0.43 & 0.40 &  0.40 &  1  &  0.42 (3.36) &   \\ 
  \hline
    \end{tabular}
    \caption{Information about the eclipsing binary candidates identified in the present study. The Table shows the internal \emph{ID} for each star. \emph{HSP} indicates their ID in \citetalias{hend2011+}, \emph{J}, \emph{H}, and \emph{K} indicates their median magnitudes in each band, and the \emph{ptp} fields show their peak to peak amplitude in each band. \emph{Type} gives their morphological classification: 1 if periodic and 3 if occultation. \emph{Per} gives the period detected for the eclipse, and \emph{Per}$_{rot}$ gives the rotational period, when measured.}
    \label{tab:EB}
\end{table*}

\begin{figure*}[tbh]
  \begin{minipage}[c]{0.75\textwidth}
    \centering
    \includegraphics[width=13cm]{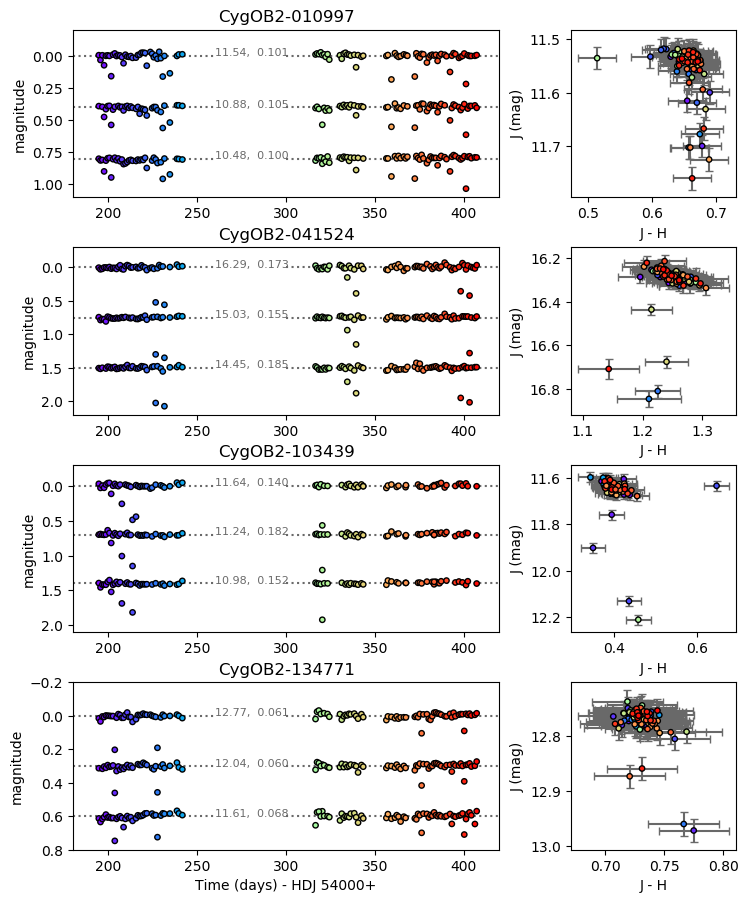}
    \end{minipage}\hfill
      \begin{minipage}[c]{0.24\textwidth}
  \caption{Four eclipsing binary candidates that were not periodic in our study. \emph{Left:} J, H, and K light curves for each star. The median magnitude and ptp amplitude are written in the plot. \emph{Right:} J vs. J-H CMDs. In both left and right plots, colors reflect the date of the observation as in Fig. \ref{fig:093173}.}
    \label{fig:EBnonP}
      \end{minipage}
\end{figure*}

Following these three criteria, we identified 14 eclipsing binary candidates. Their properties are presented in Table \ref{tab:EB}. To sum up: the identified eclipsing binary candidates presented magnitudes between 10.45 and 16.29 mag in the J band, and ptp amplitudes between 0.06 and 0.33 mag in the same band. None of them were disk-bearing stars or had measured slopes of variability in color. Their periods ranged from 0.83 to 7.30 days.

Among the eclipsing binary candidates identified, the four stars in Fig. \ref{fig:EBnonP} presented recurrent eclipses over the observed time, but these eclipses were not periodic inside the observational window of our data. 

As mentioned in the Sec. \ref{sec:sec:ec}, even though we could detect periodicity in some of the occultation variables, we often could not recover their correct period. This is partially due to the limitations of the classical implementation of the Lomb-Scargle periodogram, which searches for pure sinusoidal signals, and partially due to the observational cadence of our observations, which was not ideal for detecting periods below two days, as further discussed in \citetalias{2017Roquette}. To complement the study of the eclipsing binary candidates by taking a closer look into their periodicity, we applied two additional period search techniques alternative to the Lomb-Scargle periodogram: the fast-$\chi^2$ method \citep{2009Palmer}\footnote{Available in Python as part of the astropy.timeseries: \url{https://docs.astropy.org/en/stable/timeseries/index.html}}, which is a higher order Fourier method capable of dealing with waveforms that are more complex than a pure sinewave; and the String-Length method \citep{Clarke2002}, which is a nonparametric period search method that does not use any assumption regarding the shape of the light curve. When these two methods lead to different periods, or when we found distinct periods in the different observed bands, we visually inspected the folded light curves and their phase dispersion and based the choice of the adopted period on it. 

\begin{figure*}[tbh]
    \centering
    \includegraphics[width=16cm]{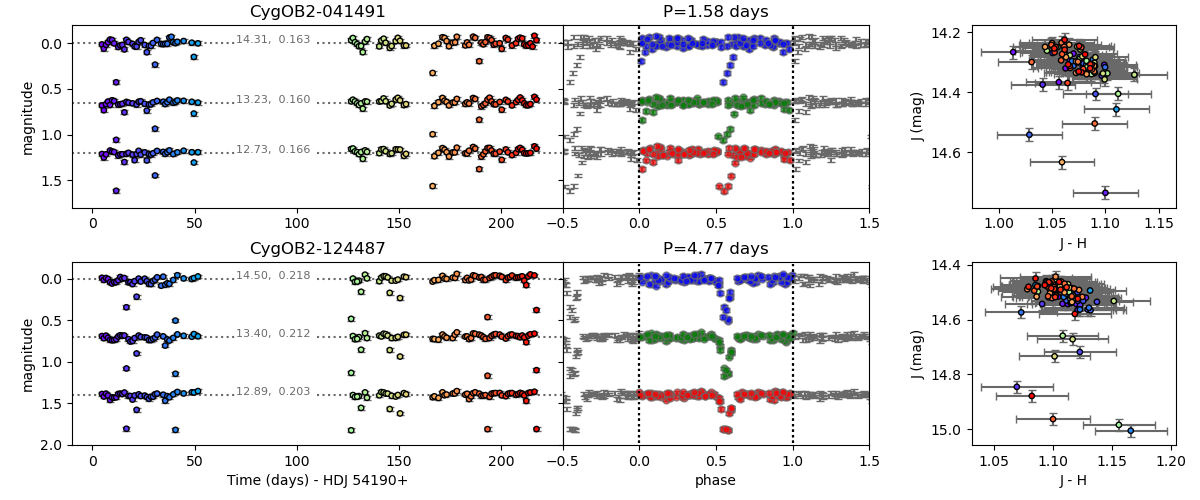}
    \caption{Two multiperiodic eclipsing binary candidates. \emph{Left:} J, H, and K light curves for each star. The median magnitude and ptp amplitude are written in the plot, and the colors reflect the date of the observation, as in Fig. \ref{fig:093173}. \emph{Middle:} phased light curve for each band, folded using the detected period attributed to the occultation feature. J, H and K bands are shown as blue, green, and red respectively.  Fig. \ref{fig:EB_multiP_2} shows the folded light curves for the secondary period. \emph{Right:} J vs. J-H CMD. Colors reflect the date of the observation, as in Fig. \ref{fig:093173}.}
    \label{fig:EB_multiP}
\end{figure*}

\begin{figure}[tbh]
    \centering
    \includegraphics[width=9cm]{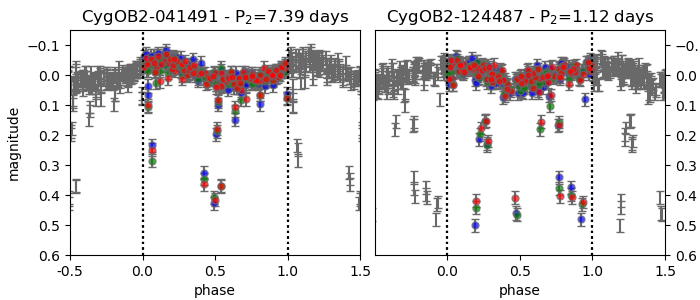}
    \caption{Phased light curve for each band, folded using the secondary detected period found for the eclipsing binary candidates presented in Fig. \ref{fig:EB_multiP}. J, H and K bands are shown as blue, green, and red respectively. }
    \label{fig:EB_multiP_2}
\end{figure}

For the stars CygOB2-041491 and CygOB2-124487, we verified the existence of two periods in the light curve. Fig. \ref{fig:EB_multiP} shows their light curves folded to the period found for the occultation features.  Fig. \ref{fig:EB_multiP_2} shows the folded light curves for the secondary period, and in both cases, the secondary period seems to be reflecting the rotational modulation of spots in the primary star in the eclipsing binary system. We remark that in Table \ref{tab:EB}, we presented the star CygOB2-041491 with two possible primary periods because different period-search techniques gave different periods for the occultation features. The period adopted for building the plot in Fig. \ref{fig:EB_multiP} was the one that minimized the phase dispersion of the occultation features. 

\begin{figure*}[tbh]
  \begin{minipage}[c]{0.7\textwidth}
    \centering
    \includegraphics[width=12cm]{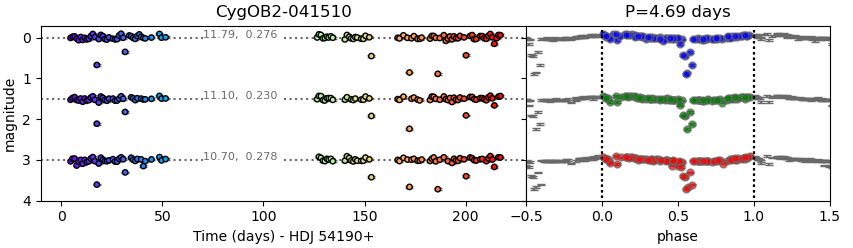}
    \end{minipage}\hfill
      \begin{minipage}[c]{0.29\textwidth}
    \caption{Eclipsing binary candidate CygOB2-041510 (Fig. \ref{fig:041510}). \emph{Left:} J, H, and K light curves for the star. The median magnitude and ptp amplitude are written in the plot, and the colors reflect the date of the observation, as in Fig. \ref{fig:093173}. \emph{Right:} phased light curve for each band, folded using the detected period attributed to the occultation feature. J, H and K bands are shown as blue, green, and red respectively.}
    \label{fig:EB041510}
      \end{minipage}
\end{figure*}

\begin{figure*}[tbh]
    \centering
    \includegraphics[width=16cm]{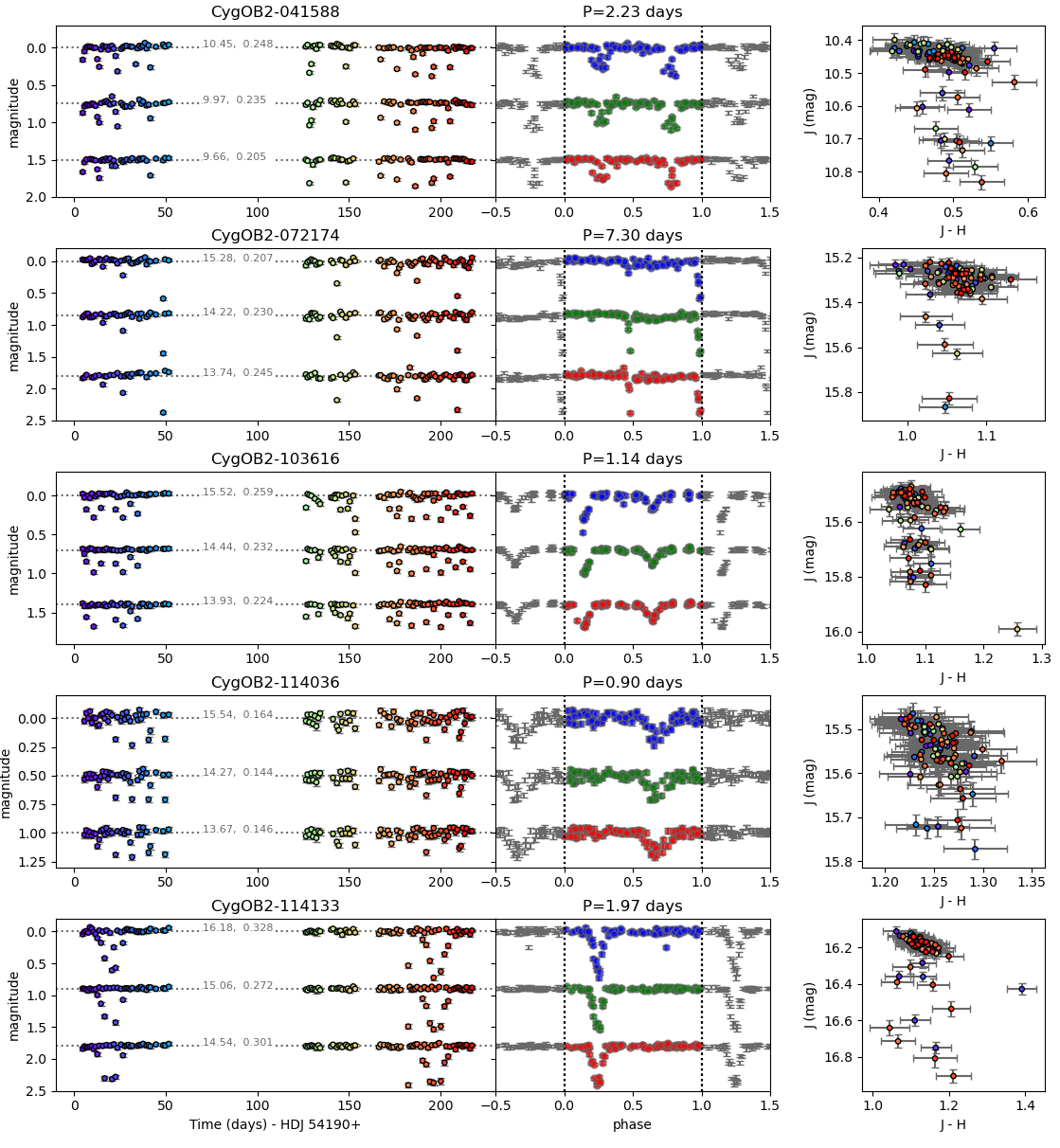}
    \caption{Five periodic eclipsing binary candidates. \emph{Left:} J, H, and K light curves for each star. The median magnitude and ptp amplitude are written in the plot, and the colors reflect the date of the observation, as in Fig. \ref{fig:093173}. \emph{Middle:} phased light curve for each band, folded using the detected period attributed to the occultation feature. J, H and K bands are shown as blue, green, and red respectively. \emph{Right:} J vs. J-H CMD. Colors reflect the date of the observation, as in Fig. \ref{fig:093173}.}
    \label{fig:singleP}
\end{figure*}

\begin{figure*}[tbh]
    \centering
    \includegraphics[width=16cm]{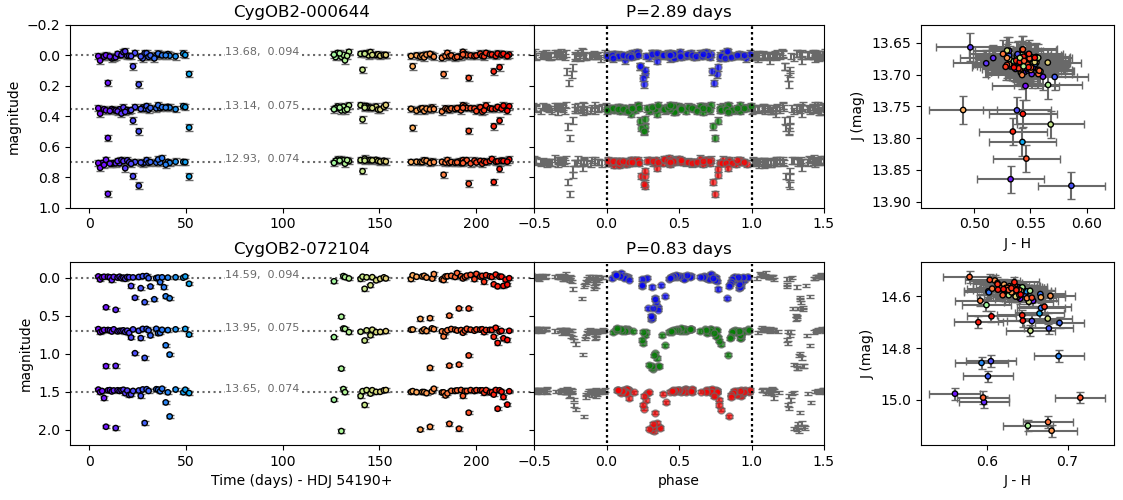}
    \caption{\label{fig:HSP11_EB} Stars CygOB2-000644 and CygOB2-072104 are both occultation and compound variables and were considered as eclipsing binary candidates in the present study. They are also reported as eclipsing binaries in HSP11. \emph{Left:} J, H, and K light curves for each star. The median magnitude and ptp amplitude are also written in the plot, and the colors reflect the date of the observation, as in Fig. \ref{fig:093173}. \emph{Middle:} phased light curve for each band. J, H and K bands are shown as blue, green, and red respectively. \emph{Right:} J vs. J-H CMD for the same star. Colors also reflect the date of the observation, as in Fig. \ref{fig:093173}.}
\end{figure*}

Fig. \ref{fig:EB041510} shows complementary plots for the eclipsing binary candidate CygOB2-041510, already presented in Fig. \ref{fig:041510}, and which has a period of 4.69 days. Seven other eclipsing binary candidates also had a single period measured and five of them are presented in Fig. \ref{fig:singleP} and two of them in Fig. \ref{fig:HSP11_EB}.

\begin{figure*}[tbh]
    \centering
    \includegraphics[width=16.5cm]{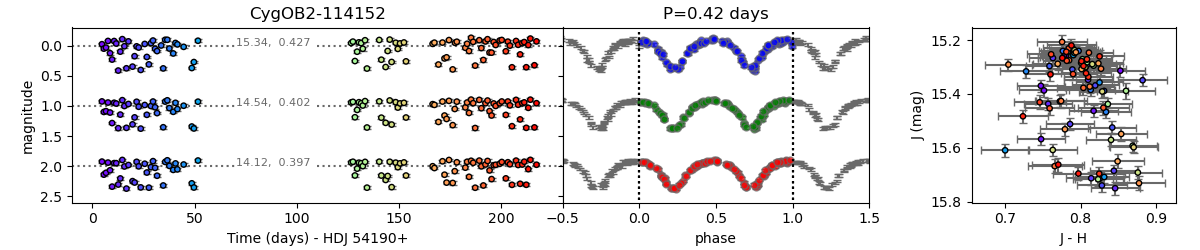}\caption{\label{fig:HSP11_Per} Star CygOB2-114152 is a periodic candidate and compound in this survey but it is reported as an eclipsing binary in HSP11. \emph{Left:} J,H, and K light curve. The median magnitude and ptp amplitudes are written in the plot, and the colors reflect the date of the observation, as in Fig. \ref{fig:093173}. \emph{Middle:} phased light curve for each band. J, H and K bands are shown as blue, green, and red respectively. The period reported by HSP11 (below our detection limit) was adopted in the phased light curve. \emph{Right:} J vs. J-H CMD. Colors reflect the date of the observation, as in Fig. \ref{fig:093173}.}
\end{figure*}

Finally, we also took into consideration the list of eclipsing binaries reported by \citetalias{hend2011+}. Among \citetalias{hend2011+} sources, we recovered in our sample four out of ten eclipsing binaries and 13 out of 17 candidates. One of their eclipsing binaries, the star CygOB2-031292, was not considered a variable within the limits of our data. Two of their eclipsing binaries were also considered as eclipsing binaries by us, and they are shown in Fig. \ref{fig:HSP11_EB}. We found a 3.21 day period for CygOB-000644 based on the classical Lomb-Scargle periodogram and on the string-length method, but a 1.447 day-period based on the fast-$\chi^2$ method. The shorter period is a harmonic of the 2.89 days period found by \citetalias{hend2011+} for the same star, which we considered as the correct period since it looks better in terms of the phase dispersion in the folded light curve. In the case of the star CygOB2-072104, we found the same 0.83 days period as in \citetalias{hend2011+}. The forth \citetalias{hend2011+} eclipsing binary in our data was the star CygOB2-114152, shown in Fig. \ref{fig:HSP11_Per}. Even though this star presented a compound behavior in the CMDs, similar to other identified eclipsing binaries, it was identified as a periodic star, but it was eliminated from the list of stars with a measured rotation period in \citetalias{2017Roquette} because it presented a different period from the one in \citetalias{hend2011+}. We found a period of 3.63 days based on the Lomb-Scargle periodogram and a 0.78 days period based on the fast-$\chi^2$ method. Both values are aliases of the 0.42330 days period reported by \citetalias{hend2011+} for the star, but the correct period is below the limit of detection of our observations. Out of the 13 eclipsing binary candidates from \citetalias{hend2011+}, in our survey two were not considered as variable within the limits of our dataset, and the 11 remaining were reported as periodic variables, but none of them presented other characteristics indicating they could be eclipsing binaries. The differences in classification are probably due to the different observational windows in the two surveys, and to account for the possibility those are indeed eclipsing binaries, in \citetalias{2017Roquette} we removed those stars from our sample of spotted stars. 

\subsection{Near-infrared variability as a disk indicator}
\label{sec:diskcandidates}

Finally, we address the reliability of using near-IR variability as a disk diagnosis. We start by looking at what are the typical variability behavior of the disk-bearing stars in our survey, and if we could use any of their typical behavior as a disk diagnosis. 

\emph{Variability amplitude:} Fig. \ref{fig:ampDfrac} shows that the disk fraction sharply increases with variability amplitude, suggesting that large amplitudes might indicate the presence of a disk. In our sample, 196 stars have ptp amplitude larger than 0.5 mag in at least one of the near-IR bands, 90$\%$ of which are disk-bearing. Out of the 12 diskless stars with such large amplitudes, 8 have slopes measured in Section \ref{sec:corrcor}. Their slopes are larger than $\gtrsim$70$^o$ in the J vs. J-H CMD, or shallower than $\lesssim$35$^o$ in the color-color diagram, implying that their variability is either disk-variability or extinction by circumstellar material, and hinting they are disk-bearing stars. 

\emph{Slopes in the color-space:} The measurement of the slopes of the trajectories in the color diagrams is a promising disk indicator as 84.8$\%$ of all slopes we measured are in known disk-bearing stars. As discussed in Sec. \ref{sec:var}, only disk variability can produce negative slopes, thus measuring those can lead to identifying disk-bearing stars. We found 115 stars with negative slopes: 110 known disk-bearing stars, and 10 newly discovered disk-bearing sources. Positive slopes can also be an indirect disk-indicator, as among their causes there are hots pots produced in the accretion process and circumstellar extinction. However, that variability caused by cold spots is independent of the presence of a disk and can also explain trajectories with positive slopes having a range of slopes that is degenerate with the ones related to disk variability. Therefore, positive slopes do not immediately imply the existence of a disk. The overlapping of the expected range of slopes for cold spots and other disk related variability is almost complete in the J vs. J-H CMD, but only partial in the K vs. H-K CMD. Therefore, it is still be possible to differentiate between disk-variability and cool spots given the positive slopes they produce in the K vs. H-K CMD, as the former causes shallower slopes than the latter (see Table \ref{tab:slopetheory}). We found 196 stars with positive slopes in the CMDs, 157 of which are known disk-bearing stars. Unfortunately, all the 39 diskless stars with a positive slope in our sample had only $\frac{\Delta J}{\Delta J-H}$ measured, and further information is required to confirm if they could indeed be disk-bearing candidates.

\emph{Compound variability:} 
In Sec. \ref{sec:sec:comp} we associated asymmetries with a distinct color in occultation variables with eclipsing binary variability, and those do not suggest the presence of a disk. Variable near-IR excess indicates the presence of a disk and accordingly the 24 compound variables of this type in our sample were known disk-bearing stars. Once eclipsing binary variability is ruled out, the only variability mechanism expected in the diskless stars is due to cold spots. As we associated compound stars with a changing slope with two or more alternating variability mechanisms, the 24 compound variable of this type must host a disk. Indeed, 21 of them were known disk-bearing stars and thus the remaining 3 changing slope variables are newly discovered disk-bearing (For example, the star CyOB2-000111 shown in Fig. \ref{fig:compound}). We also found 59 compound stars showing both variable near-IR excess and changing slopes  (see, for example, the star CygOB2-021188 in Fig. \ref{fig:021188}), 3 of which are newly discovered disk-bearing stars. 

\emph{Light curve morphology:} Periodic variability is not a good indicator of the presence of a disk as it is the most common type of variability among the diskless stars (only 10.2$\%$ of the periodic stars had a known disk). For occultation variables, once eclipsing binary candidates were identified and removed, $\sim$85$\%$ of the remaining stars (83 stars) are known disk-bearing. Those are occultation variables that can be associated with circumstellar extinction (``dippers''), and therefore also related to a disk-presence. On the other hand, 15 are diskless: four are considered disk-bearing due to their changing slope compound variability, two are considered candidates due to their positive slopes of variability in the J vs. J-H CMD. The remaining nine sources are also considered disk-bearing candidates because of their light curve morphology. 

Finally, 560 (77$\%$ of them) of the other types of variables are known disk-bearing stars. Among the 166 that are diskless: 10 are identified as new disk-bearing stars due to their negative slopes in the CMD and two due to their compound variability; 28 are identified as disk-bearing candidates due to their positive slopes in the CMDs. We thus consider that ruling out both eclipsing binary candidates and periodic variables from a list of variable stars seem to be a valuable first step toward selecting disk-bearing stars.

Considering the variability characteristics described in this section, we verified that 681 previously known disk-bearing stars in CygOB2 cab be identified as disk-bearing stars based solely on their near-IR variability. We found 18 newly discovered disk-bearing stars among the list of members of Cygnus OB2, based on their negative slopes in the CMD or in their compound variability. Additionally, we found 187 stars with ambiguous variability characteristics that could be either caused by cold spots in diskless stars, or by hot spots or extinction in disk-bearing stars, but further information is needed to confirm those as disk-bearing systems.

\section{Summary and conclusions}
\label{sec:conclusions}

In this paper, we summarized and discussed the results for the first near-IR photometric variability survey focusing on the low mass population of the Cygnus OB2 association. We observed the region in 115 nights spanning 217 days ($\sim$7 months) using WFCAM/UKIRT JHK photometry. We focused our study on a list of 5083 Cygnus OB2 low mass candidate members identified by previous studies in the literature, 25$\%$ of which were listed as disk-bearing stars. While characterizing them according to their JHK variability, we verified that:

\begin{enumerate}
\item 2529 ($\sim$50$\%$) candidate members are significantly variable in the JHK-bands. 
  \begin{enumerate}
\item 1280 of them are highly variable and present large values of the Stetson variability index (Sample $S_1$:$S\!\geq\!1$);
\item 1249 are low amplitude variables and present moderate values of the Stetson variability index (Sample $S_2$: $1\!<\!S\!\leq\! 0.25$).
  \end{enumerate}
\item Given the morphology of the light curves: 
  \begin{enumerate}
\item 1679 of variable stars show some periodicity. 10$\%$ of those are disk-bearing stars, 39 (2.3$\%$) are also periodic in color, 29 (1.7$\%$) have assymetric light curves and 1256 have a period measured in \citetalias{2017Roquette}.
\item 124 variable stars show occultation variations; 71$\%$ of which are disk-bearing stars. Among these, 83 are disk-bearing stars likely associated with AA Tau variability.

\item 726 variable stars show other types of variability, including long-term and short-term nonperiodic variability, and mixed light curve morphologies. Of those, 560 (77$\%$) are disk-bearing stars. Also, 11 are burst-variables.
\end{enumerate}

\item Disk-bearing stars are more likely to be high amplitude variables. The main results supporting this statement are:
\begin{enumerate}
\item While $\sim$57$\%$ of the sample $S_1$ is composed of disk-bearing stars, only $\sim$8$\%$ of the sample $S_2$ have disks.
\item There is a steep increase in the disk fraction for higher variability amplitude in all bands. 
\end{enumerate}
\item Periodic variability is more common among diskless stars: while 88$\%$ of all variable diskless stars present periodic behavior in their light curves, only 21$\%$ of the variable disk-bearing stars show the same behavior.
\end{enumerate}

We also explored their variability in the color-space and verified that:

\begin{enumerate}
\setcounter{enumi}{4}
\item Some variable stars present correlated variability in the color space resulting in linear trajectories inside the color-color and color-magnitude diagrams. We measured the slopes of these trajectories for 335 of such stars. We compared the slopes measured with the slopes expected from a theoretical description of the physical mechanisms typically attributed as the causes of near-IR variability in PMS stars. Considering the slopes measured inside the color-magnitude diagrams, we found that overall, our sample of stars with measured slopes is composed of a mixture of disk-bearing variability and circumstellar extinction variability, with 1.7 times more stars with variability caused by extinction/spot than caused by changes in the inner disk or accretion. 
\item 149 variable stars show a compound behavior in color. Those are variables in which the color-behavior is not restricted to a linear trajectory. We observed 3 different types of compound behavior:
\begin{enumerate}
\item Occultation with distinct color, which are variables that present occultation features with a distinct color from the rest of the light curve (42 sources).
\item Variable near-IR excess (Increasing or decreasing) that causes the data points observed in different seasons to be clumped in different regions of the color-color and color magnitude diagrams (24 sources).
\item Changing slope, which are variables showing a clear change in the dominant variability mechanism, often marked by a change in the slope of variability in the color-magnitude diagrams (24 sources).
\item A mix of (b) and (c) (59 sources).
\end{enumerate}
\end{enumerate}

\begin{enumerate}
\setcounter{enumi}{6}
\item We identified 15 eclipsing binary candidates, four of them previously identified by HSP11 and two of them presenting two periods: a period due to the eclipse, and a second period probably due to the rotational modulation of spots in the primary star. We report periods for 11 of the eclipsing binary candidates.
\end{enumerate}

Characterizing the variability of young stars in the near-IR is an efficient way of identifying disk-bearing stars and that $64.5\%$ of the previously known disk-bearing stars in Cygnus OB2 were variable in the near-IR. We also verified that $53.4\%$ (681 sources) of the known disk-bearing stars in the association can be identified as disk-bearing stars based solely on their near-IR variability. Additionally, among the member candidates of CygOB2 previously identified based on their X-ray properties, we found 18 newly identified disk-bearing stars. 

\begin{acknowledgements} This study was part of JR Ph.D. thesis, which was granted by CNPq (Conselho Nacional de Desenvolvimento Científico e Tecnológico) and CAPES (Coordenação de Aperfeiçoamento de Pessoal de Nível Superior). JR has also received funding from the European Research Council (ERC) under the European Unions Horizon 2020 research and innovation programme (grant AWESoMeStars, agreement No. 682393). S.H.P.A. acknowledges financial support from CNPq, CAPES, and FAPEMIG. J.B. acknowledges the support of ANR grant 2011 Blanc SIMI5-6 020 01 Toupies: Toward understanding the spin evolution of stars (\url{http://ipag.osug.fr/Anr_Toupies/}). We also thank Laura Venuti, Francisco Maia, and Alana Souza for useful comments over the development of the work. This study has made use of NASA’s Astrophysics Data System Bibliographic Services. This research made extensive use of TOPCAT software \citet{2005Taylor}.This research made use of Astropy,\footnote{http://www.astropy.org} a community-developed core Python package for Astronomy \citep{astropy:2013, astropy:2018}. The authors also thank the referee, Scott Wolk, for his detailed comments which helped to improve this paper. The authors wish to recognize and acknowledge the very significant cultural role and reverence that the summit of Mauna Kea has always had within the indigenous Hawaiian community. We are most fortunate to have the opportunity to use observations from this mountain. \end{acknowledgements}
%

\bibliographystyle{aa}   
\bibliography{ref}

\begin{appendix} 
\section{Estimation of typical amplitudes and slopes of spot and extinction variability in the near-IR colors}
\label{app:A}

\subsection{Spot variability}\label{app:A:spot}

To investigate the near-IR variability amplitudes produced by spots in the stellar surface we used a simple model in which two blackbodies with different temperatures describe the stellar surface and the spot, and in which the limb-darkening effects are considered. In such model, the variation in magnitude caused by a spot of temperature $T_\mathrm{spot}$ at the surface of a star of photospheric temperature $T_*$ can be estimated as:

\begin{equation}
  \Delta m=-2.5\log{\Bigg(1-\frac{f}{1-\frac{\mu}{3}}\Bigg[1-\frac{B_\lambda(T_\mathrm{spot})}{B_\lambda(T_*)}\Bigg]\Bigg)},
  \label{eq:spot}
\end{equation}
\noindent where $f$ is the fraction of the stellar surface covered by spots (filling factor), $B_\lambda(T)$ is Plank Law for the radiation of a blackbody of temperature $T$ at wavelength $\lambda$, and $\mu$ is the limb-darkening coefficient. This model does not take into account the shape, position, and number of spots.

Previous studies \citep[e.g.,][]{2001Carpenter,2009Scholz} apply similar models for estimating amplitudes and slopes of spot variability in the near-infrared color space using a small set of parameters. Here, instead, we used the parameters retrieved from observations of spotted stars. We estimated $\Delta m$ for each of the near-infrared bands by feeding the model with the parameters $T_{eff}$, $T_{spot}$ and $f$ estimated by \citet{2015Venuti} for 360 stars in NGC 2264 using optical and UV observations. We only used the parameters of stars with spectral types between F2 and M2, which covers approximately the mass interval 0.3 M$\odot$ to 1.5 M$\odot$. This same mass range contains most of the stars in our member candidate list. \citet{2015Venuti} sample includes spotted stars with $\Delta T$ (difference of temperature between the stellar photosphere and the spot) up to -6820 K for hot spots, and 7030 K for cold spots. The limb-darkening coefficients for the WFCAM JHK photometric system were provided by Dr Ant\^onio Claret\footnote{From Instituto de Astrof\'isica de Andaluc\'ia, Granada, Spain}, and were estimated as described in \citet{2011Claret}. 

For cold spots, we found typical (median) amplitudes of 0.10, 0.08, and 0.07 mag for the J, H and K bands. These values are in good agreement with the median amplitudes for the sample of diskless periodic stars analyzed in \citetalias{2017Roquette}, which were 0.10, 0.08 and 0.08 mag, supporting the idea that the periodic sample was composed mainly of stars with cold spots. On the other hand, we found that the stars with hot spots from \citet{2015Venuti} have median near-IR amplitudes of 0.07, 0.05, 0.04 mag for J, H, and K bands respectively. These values are startling low compared to the amplitudes of periods in disk-bearing sample, whose median values were 0.31, 0.30, 0.20 mag.

We showed the range of slopes of variability we found in Table \ref{tab:slopetheory} and discussed it in Sec. \ref{sec:var}. Overall we found slopes for cold and hot spot variability similar to previous studies. However, we found typical higher amplitudes for cold spots, than for hot spots, with median and max values of 0.09 and 0.61 mags in the J band for cold spots, and 0.05 and 0.5 for hot spots. The amplitudes we found were systematically lower than in previous studies for both cold and hot spots. The explanation for that is that among the spotted stars from \citet{2015Venuti}\footnote{We encourage the interested reader to refer to \citet{2015Venuti} Section 3.4, and in particular to their Fig. 10.} there is a lack of hot spotted stars with both a very hot spot and a large spot coverage. Thereby, the hypothetical star producing the max values of typical amplitudes for spotted stars in \citet{2001Carpenter} - a spotted star with a photospheric temperature of 4000 K, a spot temperature of 8000 K, and spot coverage of 30$\%$ - is not possible within the observational scenario evaluated by \citet{2015Venuti}.

It may be the case that hot spotted stars with large $\Delta T$ and spot coverage are more common than reflected by \citet{2015Venuti} sample of spotted stars, and in this case, \citet{2001Carpenter} and \citet{2009Scholz} models and parameters are reflecting more realistic amplitude ranges for this spot variability than our results. If this is not the case, then the sample of disk-bearing periodic stars presented in \citetalias{2017Roquette} cannot be composed only of hot spotted stars, and the observed variations may be caused instead by other types of periodic variability mechanisms such as variable circumstellar extinction. Either way, the spot model used and the set of parameters estimated by \citet{2015Venuti} show that there are sets of observed spot temperature and spot coverage that result in cold spot variability amplitude as high as those produced by hot spots. Hence, it is not possible to differentiate hot spots from cold spots in the near-IR based only in their variability amplitudes. It is also not possible to distinguish between hot and cold spots using their slopes, and Table \ref{tab:slopetheory} shows a significant overlap for the typical slopes in the two cases.

\subsection{Extinction variability}\label{app:A:ext}
  The wavelength dependence of extinction in the near-IR and submillimeter can be described as a power law: A$_\lambda\sim\lambda^{-\beta}$ \citep{2009Scholz}, where $\beta$ is the dust opacity index and represents the efficiency at which dust grains radiate at long wavelengths \citep[e.g.,][]{2016Sadavoy}. A typical value for $\beta$ in the interstellar medium in the IR is $\beta\sim$1.7 \citep{1990Mathis}. $\beta$ can vary even inside the same cloud \citep{2005Froebrich}, as it is a function of the size of the grain producing the opacity, and it can evolve with both density and temperature. In circumstellar disks, a typical value is $\beta\sim$1 \citep{1991Beckwith,1993Miyake,2011Williams}, but it can present values as low as 0.4 and as high as 1.6 \citep[e.g.,][]{2006Rodmann,2008Pinte,2017Garufi}. We used values in this range to estimate the slopes described by variable extinction in the CMD (Table \ref{tab:slopetheory} and Fig. \ref{fig:slopes}). Due to the nature of this type of variability, it can assume arbitrary amplitudes.
\end{appendix}

\end{document}